\documentclass{ieeeaccess}

\usepackage[hyphens]{url}
\usepackage[hidelinks]{hyperref}
\hypersetup{breaklinks=true}
\usepackage{cite}
\usepackage{amsmath,amssymb,amsfonts}
\usepackage{algorithmic}
\usepackage{graphicx}
\usepackage{textcomp}
\usepackage{afterpage}
\usepackage{array}
\usepackage{supertabular}
\usepackage{multirow}
\usepackage{multicol}
\usepackage{multirow}
\usepackage{afterpage}
\usepackage{siunitx}
\usepackage{color}
\usepackage{subfigure}
\usepackage{makecell}
\usepackage{txfonts}

\def\BibTeX{{\rm B\kern-.05em{\sc i\kern-.025em b}\kern-.08em
    T\kern-.1667em\lower.7ex\hbox{E}\kern-.125emX}}
\begin{document}
\history{Date of publication xxxx 00, 0000, date of current version xxxx 00, 0000.}
\doi{10.1109/ACCESS.2023.DOI}

\title{Low-Cost GNSS Simulators with Wireless Clock Synchronization for Indoor Positioning
}
\author{\uppercase{Woohyun Kim
\uppercase{and Jiwon Seo}}, \IEEEmembership{Member, IEEE}}
\address{School of Integrated Technology, Yonsei University, Incheon 21983, Republic of Korea}
\tfootnote{
This work was supported in part by the Institute of Information \& Communications Technology Planning \& Evaluation (IITP) funded by the Korean government (KNPA) under Grant 2019-0-01291; in part by the Future Space Navigation and Satellite Research Center through the National Research Foundation of Korea (NRF) funded by the Ministry of Science and ICT (MSIT), Republic of Korea, under Grant 2022M1A3C2074404; in part by the Unmanned Vehicles Core Technology Research and Development Program through the NRF and the Unmanned Vehicle Advanced Research Center (UVARC) funded by the MSIT, Republic of Korea, under Grant 2020M3C1C1A01086407.
}

\markboth
{Kim and Seo: Low-Cost GNSS Simulators with Wireless Clock Synchronization}
{Kim and Seo: Low-Cost GNSS Simulators with Wireless Clock Synchronization}

\corresp{Corresponding author: Jiwon Seo (jiwon.seo@yonsei.ac.kr)}

\begin{abstract}
In regions where global navigation satellite systems (GNSS) signals are unavailable, such as underground areas and tunnels,  GNSS simulators can be deployed for transmitting simulated GNSS signals.
Then, a GNSS receiver in the simulator coverage outputs the position based on the received GNSS signals (e.g., Global Positioning System (GPS) L1 signals in this study) transmitted by the corresponding simulator.
This approach provides periodic position updates to GNSS users while deploying a small number of simulators without modifying the hardware and software of user receivers. 
However, the simulator clock should be synchronized to the GNSS satellite clock to generate almost identical signals to the live-sky GNSS signals, which is necessary for seamless indoor and outdoor positioning handover. 
The conventional clock synchronization method based on the wired connection between each simulator and an outdoor GNSS antenna causes practical difficulty and increases the cost of deploying the simulators. 
This study proposes a wireless clock synchronization method based on a private time server and time delay calibration. 
Additionally, we derived the constraints for determining the optimal simulator coverage and separation between adjacent simulators.
The positioning performance of the proposed GPS simulator-based indoor positioning system was demonstrated in the underground testbed for a driving vehicle with a GPS receiver and a pedestrian with a smartphone.
The average position errors were 3.7 m for the vehicle and 9.6 m for the pedestrian during the field tests with successful indoor and outdoor positioning handovers. 
Since those errors are within the coverage of each deployed simulator, it is confirmed that the proposed system with wireless clock synchronization can effectively provide periodic position updates to users where live-sky GNSS signals are unavailable.

\end{abstract}


\titlepgskip=-15pt

\maketitle

\section{Introduction}
\label{sec:Intro}

Global navigation satellite systems (GNSS)  \cite{Sun21:Markov, Seo2009, Chen20101932, Yoon14:Medium, Park22:Horizontal, Lee22:Optimal}, including the Global Positioning System (GPS) of the United States and Galileo of Europe, provide users with accurate position and time information using satellite-based navigation signals that can be received easily almost anywhere in an open-sky environment. 
Such high accuracy and wide availability have made GNSS the most widely-used positioning, navigation, and timing (PNT) system.
However, as GNSS navigation signals are transmitted from satellites, the received signal strength on the ground is very weak. 
Consequently, GNSS is vulnerable to signal blockage and radio frequency interference (RFI) \cite{Strode16:GNSS, Hsu17:GNSS, Lee17:Monitoring, Park2021919, Park2018387, Zidan20:GNSS, Kim21:GPS}. 
Especially, receiving GNSS signals in indoor environments such as buildings, underground areas, and tunnels, is challenging \cite{Dedes05:Indoor, Diggelen02:Indoor, Groves11:Shadow}.

\begin{figure*}
  \centering
  \includegraphics[width=1.0\linewidth]{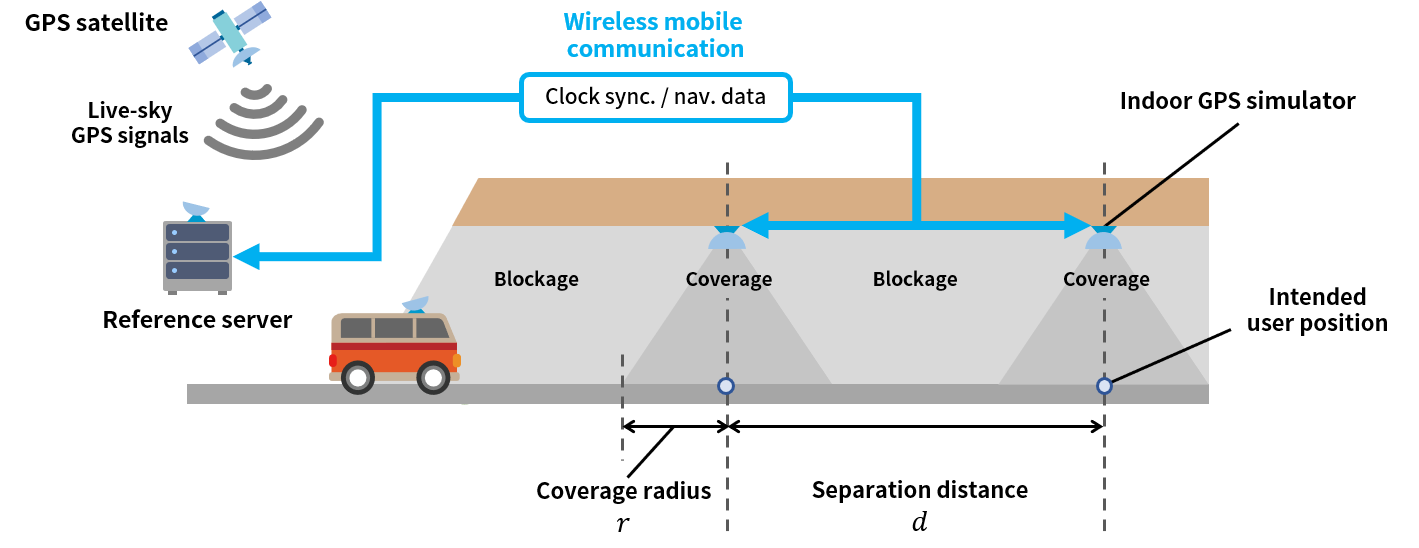}
  \caption{Overview of the proposed GPS simulator-based indoor positioning system that is capable of seamless indoor and outdoor positioning handover. The wireless clock synchronization feature between the reference server and indoor GPS simulators significantly reduces the difficulty and cost of simulator deployment.}
  \label{fig:SysOverview}
\end{figure*}

For navigating these environments, various approaches using other sensors have been studied. 
The inertial navigation system (INS) is one of the most widely-used navigation systems for both indoor and outdoor applications \cite{Ruiz12:Accurate, Cheng14:Seamless, Yadav19:Trusted, Qi02:Direct, Park2020800, Kang21:Indoor, Wen20:Multi}. 
Other commonly-used sensors for navigation include vision sensors \cite{Gaspar00:Vision, Sinopoli01:Vision, Morar20:Comprehensive, Lee18:Visual, Kim20141431, Bernardes20:Three}, ultrasonic sensors \cite{Yayan15:Low, Kapoor16:Novel, Khan21:Analysis, Rhee2019}, lidar \cite{Zou22:Comparative, Xu19:Indoor, Filipenko18:Comparison, Kim2013762}, and ultra-wide band (UWB) radar \cite{Alarifi16:Ultra, Segura12:Ultra, Elsanhoury22:Precision, Shin2017617, Lee2018:Simulation, Lee20181}. 
However, for these approaches, users must carry additional sensors other than a GNSS receiver.

To avoid the requirement of additional sensors, infrastructure-based approaches have also been studied. 
If infrastructures transmit GNSS-like signals in an indoor environment where GNSS signals are unavailable, nearby users would be able to obtain their positions with GNSS receivers. 
GNSS repeaters, pseudolites, or GNSS simulators can serve as the infrastructure for this purpose.

A GNSS repeater \cite{Jardak09:Indoor, Fluerasu09:GNSS, Im06:Indoor} broadcasts the GNSS signals received by an outdoor antenna through an indoor antenna. 
Within the coverage of a repeater, a user's GNSS receiver acquires delayed signals proportional to the distance between the repeater's indoor transmitting antenna and GNSS receiver's receiving antenna, resulting in a uniform increase in the pseudorange measurements across all visible GNSS satellites.
As a result, the position output of the GNSS receiver is limited to the position of the repeater's outdoor antenna but the receiver's time correction will be changing according to the distance to the repeater's indoor antenna \cite{Zalewski14:Real-time}.
Delay-based continuous positioning using multiple indoor repeater antennas \cite{Jardak09:Indoor, Fluerasu09:GNSS} has also been studied, but a GNSS receiver must have prior knowledge of the repeater's antenna positions. 
Thus, software modifications of GNSS receivers in the market are required to use this technique.  

A pseudolite \cite{Wang02:Pseudolite, Kee03:Indoor, Kim14:Pseudolite, Gan19:New} transmits specially designed navigation signals from a fixed location.
If a user receiver can receive and process such signals from four or more pseudolites, the receiver can calculate its 3D position similar to the GNSS positioning. 
However, a specially designed receiver may be required to process the pseudolite signals, and many pseudolites would need to be deployed to provide a sufficient geometric distribution of transmitters for users in large and complex indoor spaces.

A GNSS simulator \cite{Xu17:Improved, Perdue15:Zone, Kim19:Effect, Kim20181087} generates simulated GNSS signals so indoor user receivers can output the intended position by solving it via reception of these signals. 
In this case, unmodified conventional GNSS receivers should be sufficient to generate the position output. 
However, as discussed in the GNSS repeater case, user receivers at different locations within the same coverage of a certain GNSS simulator display the same position but varying receiver time corrections according to the distance between the receiver and simulator antennas \cite{Zalewski14:Real-time}. 
Unlike the pseudolite case, the user position cannot be continuously updated; however, it is updated only when the user moves from one simulator coverage to another simulator coverage.
Nevertheless, the GNSS simulator approach requires fewer transmitters than the pseudolite approach because a user receiver must receive only one signal instead of a minimum of four transmitter signals. 

A major technical challenge to implement such an indoor positioning system using GPS simulators, which is the topic of this study, is seamless positioning handover between the live-sky GPS signals and simulated indoor GPS signals.
As illustrated in Fig. \ref{fig:SysOverview}, live-sky GPS signals are blocked when a car enters a tunnel. 
When the car enters the coverage of the first GPS simulator, the GPS receiver of the car should be able to acquire and track the simulated GPS signals and output the simulated position before it exits the coverage.
The simulated position (that is, intended user position) from each GPS simulator can be prearranged as the 3D position of the center point of the simulator coverage.

For the receiver to quickly acquire and track the simulated GPS signals after the brief signal blockage and output the appropriate position solution, the simulated GPS signals should be very similar to the live-sky GPS signals. 
Specifically, the clock of the GPS simulator should be synchronized with the GPS satellite clock, and the navigation messages modulated on the simulated signals should be the same as those on the live-sky signals. 
Otherwise, the receiver may not output the desired position before the car passes through the simulator coverage. 

Commercial off-the-shelf (COTS) GNSS simulators support clock synchronization via pulse-per-second (PPS) inputs. 
For example, in \cite{Perdue15:Zone}, the clock synchronization of GPS simulators was achieved using PPS signals from commercial high-precision clock synchronization equipment connected to the GPS simulators via wires. 
However, clock synchronization based on wired connections increases the difficulty and cost of simulator deployment in actual indoor environments. 
Usually, PPS signals are generated by a GNSS timing receiver connected to an outdoor GNSS antenna.
The wiring between the outdoor antenna and indoor timing receiver can be very complex, especially when deployed in a tunnel or underground area. 
Considering the practicality of GNSS simulator deployment, wireless clock synchronization is much more appealing, although wireless clock synchronization is less precise than the wired case and challenging to achieve due to the uncertainty of the connection \cite{Wu11:Clock}.

Another technical issue in implementing an indoor positioning system using GPS simulators is the optimal placement of the simulators.
If the separation between adjacent simulators decreases, the number of simulators required to cover the same area increases, thus increasing the overall system cost.
If the separation between simulators increases, the signal blockage time (e.g., Fig. \ref{fig:SysOverview}) increases, and thus a user receiver may not quickly acquire and track the simulated signals after the signal blockage. 
Therefore, an optimal separation between simulators must be designed for deployment. 





This study addresses the two aforementioned technical challenges for implementing GPS simulator-based indoor positioning system. 
We propose methods for wireless clock synchronization and derive the constraints for designing the optimal simulator coverage and separation between adjacent simulators, which can significantly improve the practicality of the system deployment. 

An underground testbed was implemented to demonstrate the proposed methods' performance.
The reference server was connected to an outdoor GPS receiver that received live-sky GPS signals and decoded the real-time GPS navigation data.
The outdoor GPS receiver also provided the GPS timing information to the reference server via PPS signals.
The server and the indoor GPS simulators were connected via long-term evolution (LTE) wireless communication channels to distribute navigation data and synchronize clocks wirelessly.  
Field tests demonstrated the positioning capability of the implemented testbed for a moving car and pedestrian. 



The contributions of this study are summarized as follows:

\begin{itemize}
  \item We analyzed GPS simulator's required clock synchronization accuracy for seamless indoor and outdoor positioning handover. 
  The required level of accuracy we found opened the possibility of wirelessly synchronizing the simulator clocks. 
  \item We proposed methods for the wireless clock synchronization of GPS simulators with the required accuracy.
  A private time server was implemented, and the time delay in the GPS simulation process was measured and calibrated. 
  This wireless clock synchronization reduces the difficulty and cost of system deployment. 
  \item We derived the constraints on the simulator coverage and separation distance between adjacent simulators. 
  This information is useful for optimally deploying the proposed system. 
  \item We implemented an underground testbed with three simulators transmitting live signals to demonstrate the proposed system's positioning performance. 
  The field tests were performed for a driving vehicle with a GPS receiver and a pedestrian with a smartphone.
\end{itemize}

The remainder of this paper is organized as follows. 
Section \ref{sec:ClockErrReq} analyzes the required clock synchronization accuracy for the proposed indoor positioning system, and Section \ref{sec:ClockCorr} describes our wireless clock synchronization methods. 
Section \ref{sec:CoverageModel} derives the constraints for optimal deployment of the GPS simulators. 
Section \ref{sec:Results} presents field test results that demonstrate the performance of the proposed indoor positioning system. 
Finally, Section \ref{sec:Conclusion} concludes the study.

\section{Required Clock Synchronization Accuracy for GPS Simulator-Based Indoor Positioning System}
\label{sec:ClockErrReq}

\subsection{Importance of Clock Synchronization for Seamless Positioning Handover Between Live-Sky and Simulated GPS Signals} 
\label{sec:OpMode}

The operation of a GPS receiver channel switches between the acquisition and tracking modes \cite{Borre07:Software}. 
In the acquisition mode, coarse search is performed by changing the pseudorandom noise (PRN) code, code delay, and Doppler frequency of the local replica of the GPS signal. 
In the tracking mode, the code and carrier phases of the GPS signal are precisely tracked to generate pseudorange measurements. 
The quality of the tracking status is also evaluated.

If tracking a certain GPS satellite signal is interrupted by signal blockage or other reasons, the corresponding receiver channel returns to the acquisition mode.
Because the receiver channel already has the previously tracked GPS signal information, such as PRN code, code delay, and Doppler frequency, the channel can quickly reacquire the lost signal when the signal is reintroduced if the signal loss duration is brief. 
Reacquiring the lost signal is called reacquisition, and the time required for the receiver to output a position solution when the signal is reintroduced after the interruption is called the reacquisition time \cite{Soloviev04:Implementation, Wang14:Scheme, Kim19:Effect}.

The clock synchronization error of the GPS simulator causes a difference in code delay between the live-sky and simulated GPS signals. 
Thus, the receiver channel has slow reacquisition after a temporary blockage if the synchronization error is significant because the previous code delay information before the blockage would be very different from the current code delay information.
This is problematic because fast reacquisition is crucial for seamless positioning handover between the live-sky and simulated GPS signals. 
Furthermore, the clock synchronization error leads to a position error in a user receiver.

To develop an appropriate wireless clock synchronization method for a GPS simulator-based indoor positioning system, the clock synchronization accuracy that enables fast reacquisition and ensures acceptable positioning accuracy must be known.
Therefore, we established the requirement for the clock synchronization of a GPS simulator through performance tests of a conventional GPS receiver, which is described in Section \ref{sec:ReacqChar}.

It should be noted that the GPS simulator coverages in Fig. \ref{fig:SysOverview} do not overlap, and there are blockage regions between the simulators on purpose.
If the coverages of different simulators overlap, interference and multipath issues need to be resolved. 
As a straightforward solution, we propose non-overlapping simulator coverages. 
This configuration has additional benefit of reducing the number of deployed simulators for a given area, thus reducing the overall cost of the system.



\subsection{Analysis for Establishing Clock Synchronization Accuracy Requirement}
\label{sec:ReacqChar}

\begin{figure}
  \centering
  \includegraphics[width=0.9\linewidth]{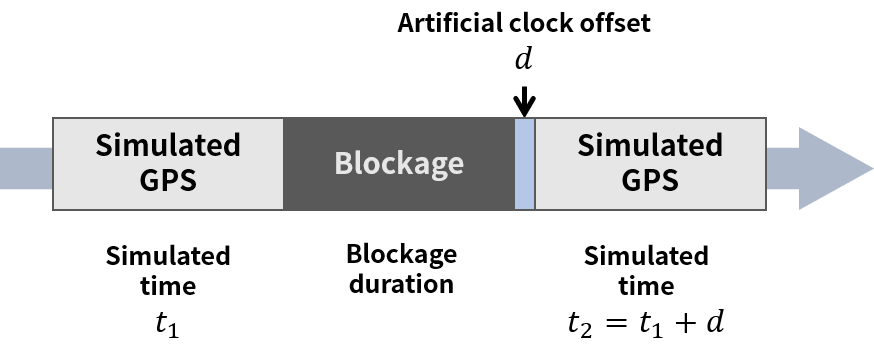}
  \caption{Types of signals provided to the GPS receiver over time in the experiment for analyzing the effect of the clock error of a GPS simulator.}
  \label{fig:ReacqTest}
\end{figure}

We designed a test to analyze the reacquisition performance of a conventional GPS receiver under simulated GPS signals according to the clock synchronization error of the GPS simulator.
Fig. \ref{fig:ReacqTest} illustrates the signal conditions provided over time to the tested GPS receiver. 
Here, we generated the simulated GPS signals before and after the blockage based on the GPS simulator clock. 
The transmission and blockage durations of the simulated GPS signals were both 60 seconds. 

The GPS simulator in this study was developed based on the open-source software, \textit{GPS-SDR-SIM} \cite{Ebinuma:GPS}. 
The \textit{GPS-SDR-SIM} uses recorded ephemeris files to generate baseband signals for offline testing, and does not support real-time GPS simulation.
For the real-time GPS simulation of this study, we modified the source codes and implemented a function to enable the transmission of signals synchronized to the current time.
In addition, we developed the functionality to reproduce the navigation data, such as telemetry (TLM) words, for the simulator according to the interface specifications \cite{GPSIS22} based on the real-time raw navigation data provided by the reference server. 
Consequently, the simulated GPS signals from our simulator become almost identical to the current live-sky GPS signals.

To simulate the clock synchronization error of the GPS simulator, we generated an artificial clock offset $d$ in Fig. \ref{fig:ReacqTest} when simulating the GPS signal after the blockage.
Unlike the previous study \cite{Kim19:Effect}, where only a positive clock offset was considered, we also tested the negative $d$ case. 
This means that the reacquisition performance of a GPS receiver was tested when the clock of the GPS simulator was slower than that of the actual GPS satellite and when it was faster. 
A \textit{u-blox M8T} GNSS receiver was utilized as a conventional GPS receiver. 
The position output rate of the receiver was set to 10 Hz. 
The receiver was connected to the GPS simulator with a cable to protect it from other possible error sources.  
The reacquisition time of the receiver was measured by changing the clock offset $d$ from $-250$ ms to $+250$ ms with an increment of 50 ms under the test scenario shown in Fig. \ref{fig:ReacqTest}. 

\begin{figure}
  \centering
  \includegraphics[width=1.0\linewidth]{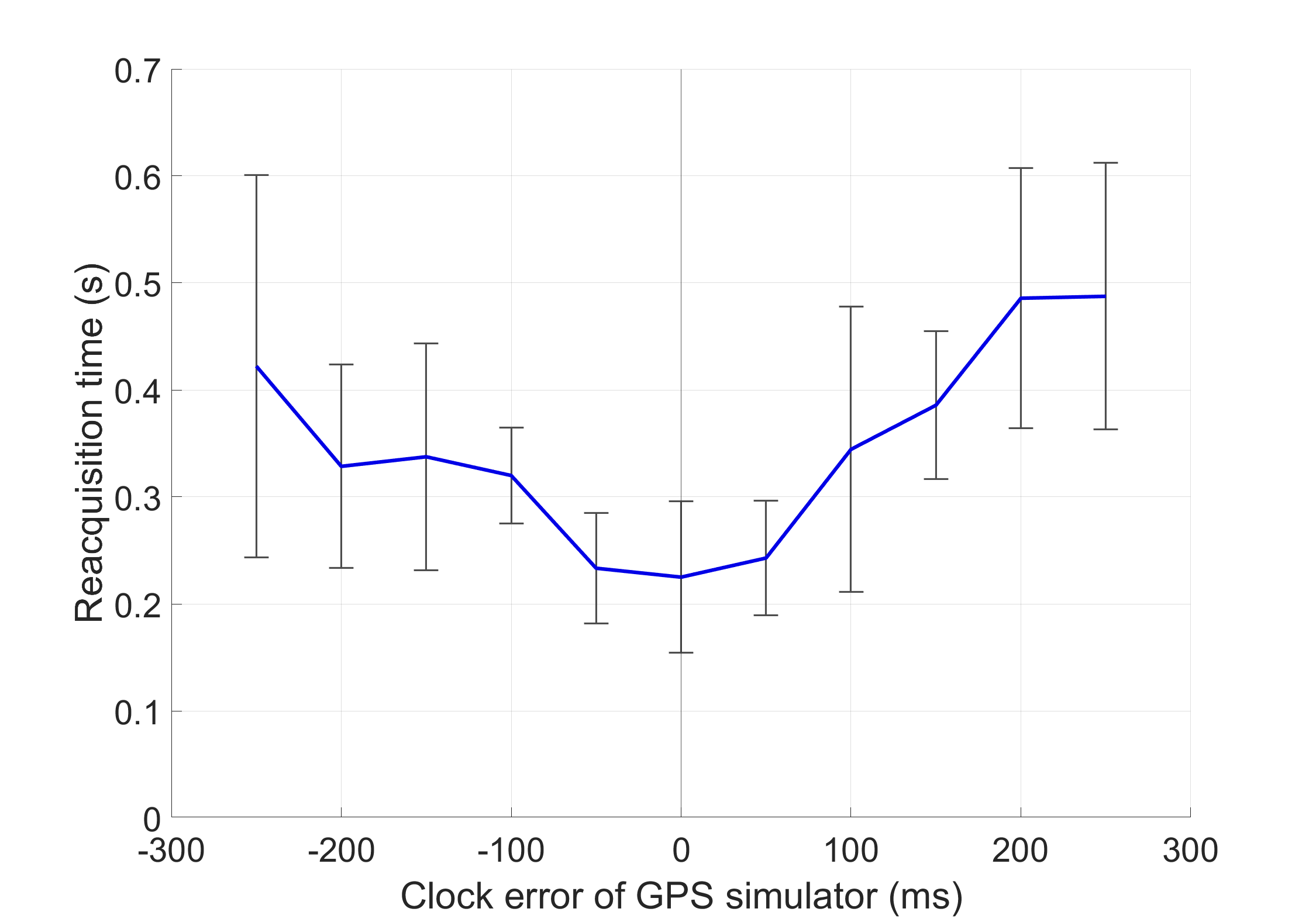}
  \caption{Measured reacquisition times of a user receiver according to the clock errors of the GPS simulator. The standard deviations of the measured reacquisition times are indicated by error bars.}
  \label{fig:ReacqTime}
\end{figure}

\begin{figure}
  \centering
  \includegraphics[width=1.0\linewidth]{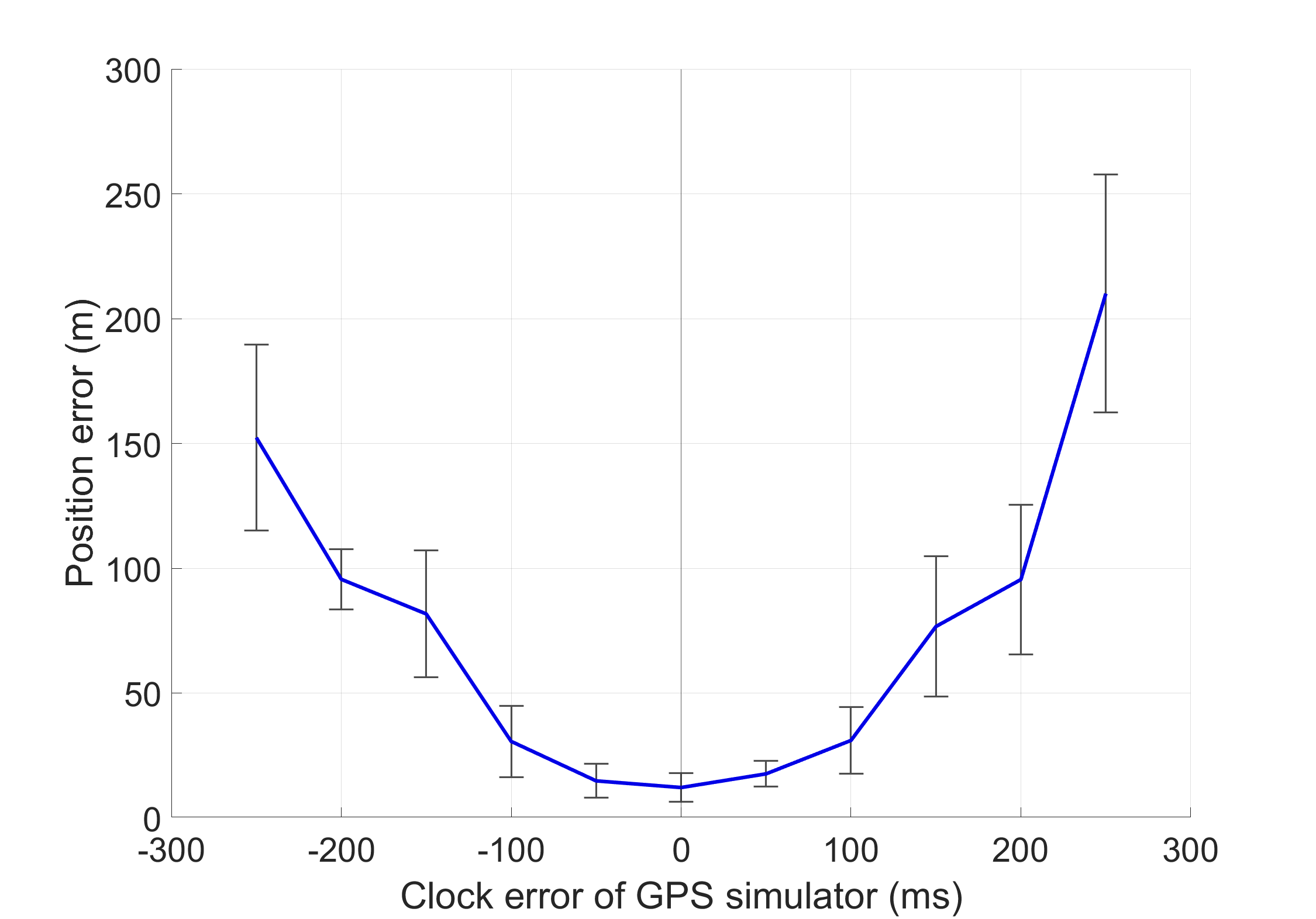}
  \caption{Measured GPS position errors of a user receiver based on the simulated GPS signals according to the clock errors of the GPS simulator.
  The standard deviations of the measured position errors are indicated by error bars.}
  \label{fig:ReacqPos}
\end{figure}

Fig. \ref{fig:ReacqTime} shows the statistics of the measured reacquisition times. 
As expected, the reacquisition time tends to increase as the simulated clock synchronization error magnitude increases. 
Furthermore, the clock synchronization error causes increased positioning error.
Fig. \ref{fig:ReacqPos} presents the statistics of the measured position errors of the user receiver according to the clock errors of the GPS simulator. 
Here, the position error is defined as the difference between the intended user position of the simulator and actual position output from the user receiver.


The required clock synchronization accuracy can be established based on these test results.
When the clock synchronization error was within 50 ms, the reacquisition time in Fig. \ref{fig:ReacqTime} and position error in Fig. \ref{fig:ReacqPos} were close to the cases with no clock synchronization error. 
This is an important finding because this level of synchronization accuracy may not require wired clock synchronization.
We propose methods to achieve a better than 50 ms clock synchronization accuracy wirelessly in Section \ref{sec:ClockCorr}.
Our wireless solution provides significant benefits over a wired solution for the cost and convenience of system deployment. 


As mentioned in Section \ref{sec:Intro}, the indoor positioning system in Fig. \ref{fig:SysOverview} does not intend to provide continuous position updates for users.
The position output from a user receiver is updated only when the receiver enters the next simulator's coverage. 
The pseudorange measurements include the time delay effect according to the distance between the receiver and simulator antennas. 
Since this effect is uniform across all simulated satellites, the receiver's position solution perceives it as a change in the receiver's time correction \cite{Zalewski14:Real-time}.
Therefore, within the same coverage, the position output from the receiver remains the same as the intended user position of the simulator (that is, the center point of its coverage), while the time correction output changes according to the distance between the receiver and simulator antennas.
Although this system cannot provide continuous position updates to a car, the periodic position update enabled by this system in a tunnel or underground area, where no GPS position update is available, will be beneficial for a car GPS navigation system to provide proper driving guidance to a destination. 
It should be noted that the user's GPS receiver does not require any hardware or software modifications to calculate its position utilizing the signals from the proposed GPS-simulator-based indoor positioning system. 



\section{Proposed Wireless Clock Synchronization Method}
\label{sec:ClockCorr}

To achieve a wireless clock synchronization of GPS simulators with an accuracy of better than 50 ms, we utilized the network time protocol (NTP) \cite{Mills91:Internet}. 
Since NTP is a protocol for synchronizing clocks through a network connection, it has low implementation difficulty and cost. 
The accuracy of clock synchronization using NTP in a wired local area network (LAN) connection is approximately 1 ms \cite{Ridoux09:Ten}.
However, the GPS simulators in our proposed system in Fig. \ref{fig:SysOverview} are connected to the reference server through a wireless LTE mobile network.
Thus, a clock synchronization error of more than 50 ms was frequently observed when we directly applied NTP in the wireless network. 
Therefore, this section proposes methods for mitigating the NTP clock synchronization error for synchronization accuracy within the 50 ms requirement.

\subsection{Clock Synchronization Error Model}
\label{sec:SyncErrModel} 

\begin{figure}
  \centering
  \includegraphics[height=7cm]{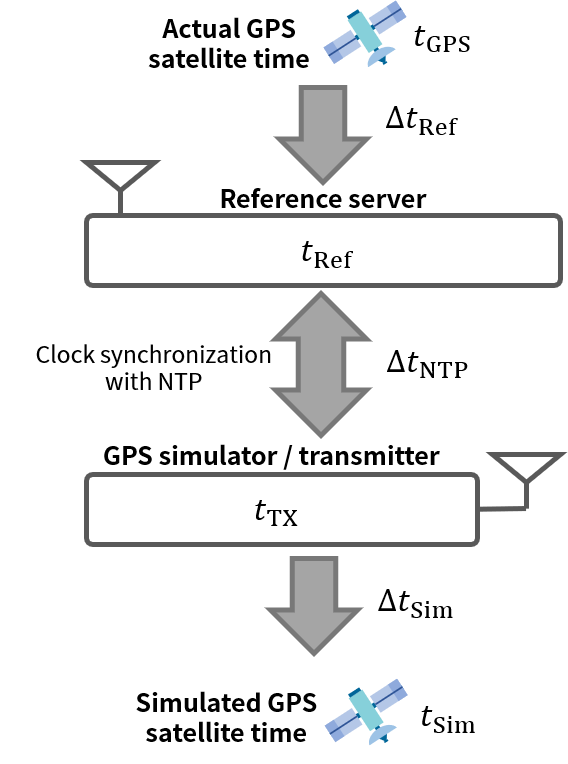}
  \caption{Clock synchronization error model of the proposed GPS simulator. $\Delta t_\textrm{NTP}$ is the most problematic error component for wireless clock synchronization over the LTE network.}
  \label{fig:SyncErrModel}
\end{figure}

The components of the clock synchronization error in the proposed system are illustrated in Fig. \ref{fig:SyncErrModel}.
The $t_\textrm{GPS}$, $t_\textrm{Ref}$, $t_\textrm{TX}$, and $t_\textrm{Sim}$ represent the time scales of an actual GPS satellite, reference server, GPS simulator, and simulated GPS satellite, respectively. 
Once a GPS receiver calculates its position and clock bias based on the simulated GPS signals, the receiver's clock can be synchronized to the simulated GPS satellite's clock. 
Thus, the time scale of a user receiver is not explicitly shown in this figure. 

The system-wide clock synchronization error (i.e., $t_\textrm{Sim} - t_\textrm{GPS}$) can be expressed as follows. 
\begin{equation}
  \label{eqn:SystemClockError}
  \begin{split}
    & t_\textrm{Sim} - t_\textrm{GPS} \\
    & = (t_\textrm{Sim} - t_\textrm{TX}) + (t_\textrm{TX} - t_\textrm{Ref}) + (t_\textrm{Ref} - t_\textrm{GPS})\\
    & = \Delta t_\textrm{Sim} + \Delta t_\textrm{NTP} + \Delta t_\textrm{Ref}
  \end{split}
\end{equation}
The clock synchronization error in \eqref{eqn:SystemClockError} should be less than 50 ms, as discussed in Section \ref{sec:ReacqChar}.


$\Delta t_\textrm{Ref}$ in \eqref{eqn:SystemClockError} is the synchronization error between the clocks of the actual GPS satellite and GPS timing receiver connected to the reference server.
$\Delta t_\textrm{Ref}$ is negligible for the 50 ms error requirement,  which will be discussed in Section \ref{sec:PrivateNTP}. 
$\Delta t_\textrm{Sim}$ is the delay that occurs during the GPS signal simulation process, which can be measured and calibrated because this delay does not significantly change over time. 
However, $\Delta t_\textrm{NTP}$ is the most problematic.
$\Delta t_\textrm{NTP}$ represents the synchronization error between the reference server and simulator clocks, which cannot be calibrated because it continuously changes depending on the status of the wireless network connection between the reference server and simulator.
Therefore, we propose methods to reduce $\Delta t_\textrm{NTP}$ using a private NTP server in Section \ref{sec:PrivateNTP} and calibrate $\Delta t_\textrm{Sim}$ using the measurement setup in Section \ref{sec:DelayCalib}.

\subsection{Wireless Clock Synchronization Using a Private NTP Server over LTE Network} 
\label{sec:PrivateNTP}

An NTP time server can be configured as sub-servers of several layers. 
The stratum number of an NTP server indicates the number of layers between the server and a high-precision time source. 
High-precision time sources such as atomic clocks or GNSS timing receivers are considered stratum 0 devices.
A stratum 1 time server is connected directly to a stratum 0 device and serves as a primary time source on the network. 
A stratum $n$ server provides time information received from a stratum $n-1$ server. 
If the stratum number exceeds 15, the clock is considered out of sync \cite{Mills06:Computer}.

\begin{figure}
  \centering
  \includegraphics[width=1.0\linewidth]{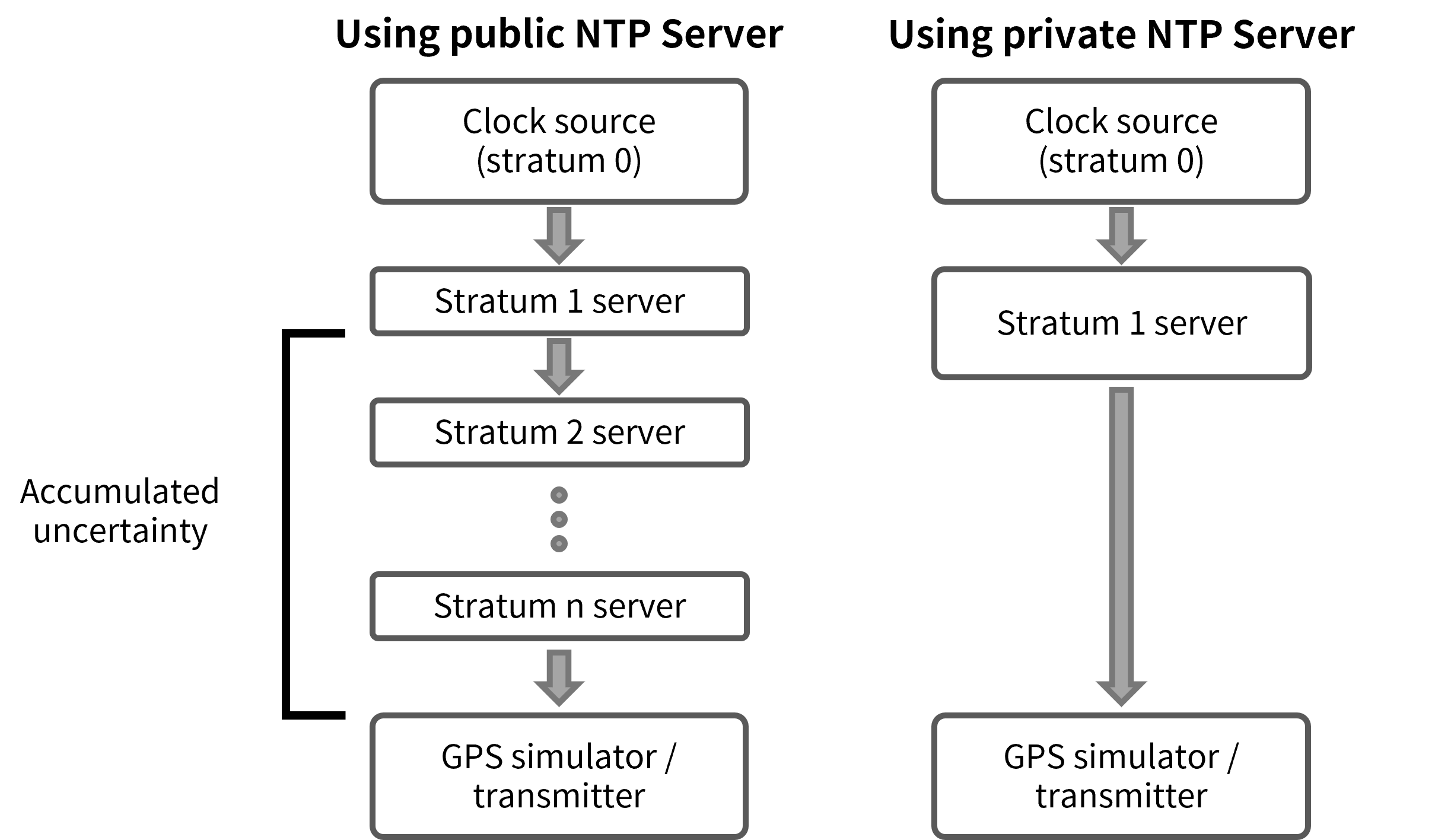}
  \caption{Comparison between the public NTP server and private NTP server cases for the clock synchronization of GPS simulators. Public NTP servers use multiple layers of secondary time servers to alleviate traffic concentration; thus, the clock uncertainty is accumulated over each layer.}
  \label{fig:PrivateNTP}
\end{figure}

In many cases, a public NTP server handles clock synchronization requests through stratum 2 or more secondary servers to disperse the traffic of clock synchronization requests. 
However, this is disadvantageous in our application because the clock error of the GPS simulator, which is an NTP client, increases as the clock errors of the secondary time servers accumulate. 
Fig. \ref{fig:PrivateNTP} illustrates the difference between a public NTP server case and a private stratum 1 NTP server case.
To minimize the clock error accumulated over the strata of the secondary time servers, we implemented a private stratum 1 time server for the direct connection to the GPS simulators.

One way to configure a stratum 1 time server is to utilize the precise PPS outputs of a reference GNSS timing receiver. 
We used a low-cost single board computer (SBC) with general purpose input/output (GPIO) pins, \textit{Raspberry Pi 3}, to receive PPS signals for implementing a private stratum 1 time server. 
As the reference GNSS timing receiver, the \textit{u-blox M8T} GNSS timing receiver was used. 
The timestamps and PPS outputs from the GNSS timing receiver were processed by an open-source NTP software, \textit{chrony} \cite{Curnow:chrony}, for synchronizing the clock of the \textit{Raspberry Pi 3} with the coordinated universal time (UTC).  
According to the maximum clock error estimated by \textit{chrony}, the clock synchronization error between the private NTP server and UTC, approximately the same as $\Delta t_\textrm{Ref}$ in \eqref{eqn:SystemClockError}, was less than 200 ns.
Although $\Delta t_\textrm{Ref}$ in \eqref{eqn:SystemClockError} is expressed in terms of the GPS time (GPST), $t_\textrm{GPS}$, UTC can be estimated based on GPST with an approximately 25 ns accuracy \cite{Enge11:global}. 
The 200 and 25 ns errors are negligible compared with our 50 ms error budget.  

\begin{table}
  \caption{Comparison of clock synchronization performance according to network connection and NTP server types.}
  \label{table:NTPPerformance}
  \centering
  \begin{tabular}{|c|c|c|}
    \hline
    Connection type & NTP Server type & Estimated maximum $\Delta t_\textrm{NTP}$ \\ \hline
    \multirow{2}{*}{\begin{tabular}[c]{@{}l@{}}Wired\\ network\end{tabular}} & Public & 15 ms \\ \cline{2-3} 
     & Private & 1 ms \\ \hline
    \multirow{2}{*}{\begin{tabular}[c]{@{}l@{}}Wireless\\ LTE network\end{tabular}} & Public & 75 ms  \\ \cline{2-3} 
     & Private & 20 ms \\ \hline
  \end{tabular}
\end{table}

A private NTP server was implemented to reduce $\Delta t_\textrm{NTP}$ in \eqref{eqn:SystemClockError}, which is the most problematic error component for wireless clock synchronization.
Table \ref{table:NTPPerformance} compares the estimated maximum $\Delta t_\textrm{NTP}$ according to the network connection and NTP server types used for the clock synchronization of the GPS simulator.
For the public NTP server in the experiment, we selected the NTP server with the lowest measured clock error among the servers in the Korean NTP pool, a collection of public NTP servers in South Korea \cite{NTPPool}. 
The estimated maximum $\Delta t_\textrm{NTP}$ values in Table \ref{table:NTPPerformance} were provided by \textit{chrony}. 
Those values do not represent $\Delta t_\textrm{NTP}$ but the estimated upper bound of $\Delta t_\textrm{NTP}$.

As shown in Table \ref{table:NTPPerformance}, the clock synchronization performance using a wireless LTE network is significantly worse than for a wired network. 
To satisfy the 50 ms error requirement of the proposed wireless GPS simulator, the private NTP server we implemented is necessary for clock synchronization.

\begin{table*}
  \caption{Comparison of the user position errors according to the clock correction methods under a static test.}
  \label{table:ClockCorrPos}
  \centering
  \begin{tabular}{|c|c|c|c|c|} 
  \hline
  \makecell{} & 
  \makecell{Public NTP server only} & 
  \makecell{Public NTP server with \\ $\Delta t_\textrm{Sim}$ calibration} &
  \makecell{Private NTP server only} & 
  \makecell{Private NTP server with \\ $\Delta t_\textrm{Sim}$ calibration} \\
  \hline
  \makecell{Maximum position error} & 65.54 m & 50.28 m & 13.26 m & 10.24 m \\ 
  \hline
  95\% position error & 11.79 m & 8.68 m & 6.16 m & 2.95 m \\ 
  \hline
  \end{tabular}
\end{table*}

\subsection{Time delay calibration in GPS simulation process}
\label{sec:DelayCalib} 

The time delay in the GPS simulation process, $\Delta t_\textrm{Sim}$ in \eqref{eqn:SystemClockError}, includes the computational delay that occurs during the generation of GPS signals based on the input navigation data and parameters, and the transmission delay that occurs during the transmission process of the generated GPS signals with the RF and antenna components. 
Since $\Delta t_\textrm{Sim}$ is related to the simulator hardware specifications, its variation over time is small for the given hardware.
Thus, $\Delta t_\textrm{Sim}$ can be mitigated by one-time calibration.

\begin{figure}
  \centering
  \includegraphics[width=0.9\linewidth]{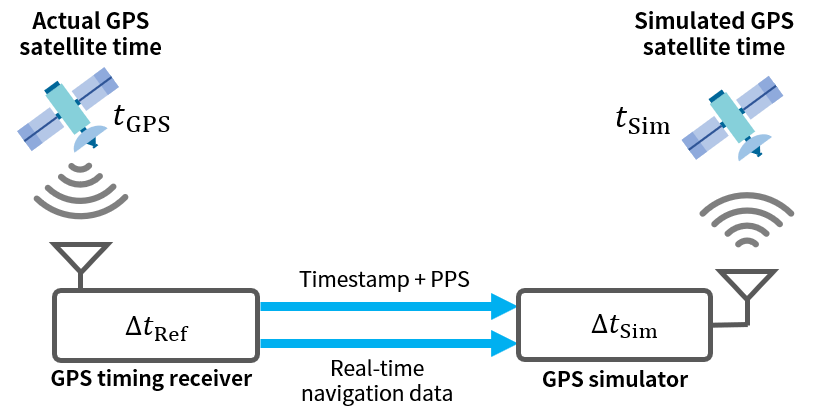}
  \caption{Measurement setup for measuring the time delay of the GPS simulation process, $\Delta t_\textrm{Sim}$.}
  \label{fig:DelayTest}
\end{figure}

To measure the $\Delta t_\textrm{Sim}$ of the given simulator hardware, we implemented a measurement setup with a GPS simulator and GPS timing receiver, as shown in Fig. \ref{fig:DelayTest}.
Here, the clock of the GPS simulator is synchronized via PPS signals with the clock of the GPS timing receiver that receives live-sky GPS signals.
As mentioned in Section \ref{sec:SyncErrModel}, the clock synchronization error caused by the GPS timing receiver, $\Delta t_\textrm{Ref}$ in \eqref{eqn:SystemClockError}, is negligible for our application because we require a 50 ms clock synchronization accuracy.
Since the GPS timing receiver is directly connected to the GPS simulator by a wire in this measurement setup, the $\Delta t_\textrm{NTP}$ error in Fig. \ref{fig:SyncErrModel} is also negligible. 
Therefore, the difference between the actual and simulated GPS satellite time in Fig. \ref{fig:DelayTest} can be considered as $\Delta t_\textrm{Sim}$.

\begin{figure}
  \centering
  \includegraphics[width=1.0\linewidth]{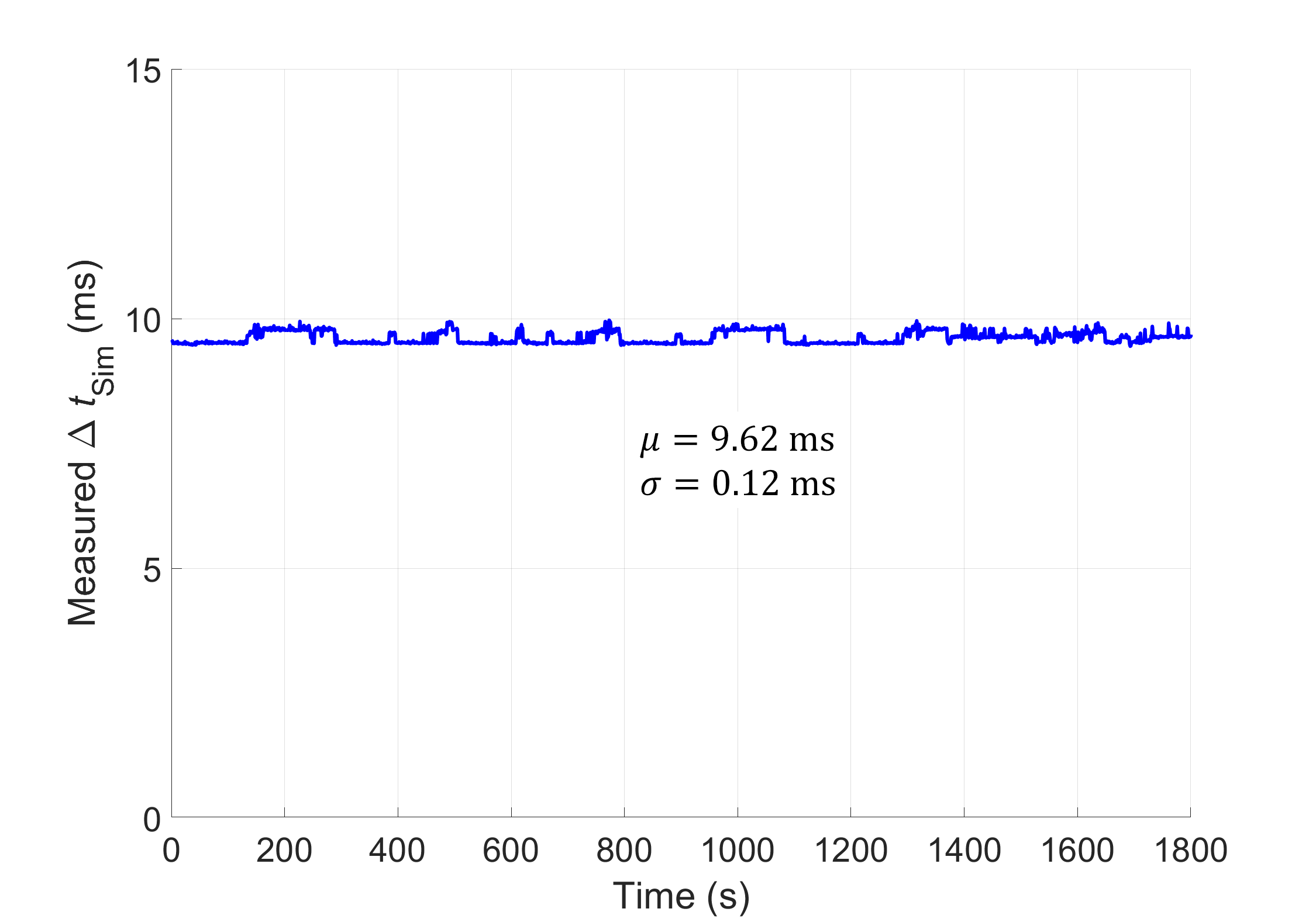}
  \caption{$\Delta t_\textrm{Sim}$ measured every second for 30 min. The mean and standard deviation of $\Delta t_\textrm{Sim}$ measurements are also shown.}
  \label{fig:TXDelay}
\end{figure}

To measure this time difference, $\Delta t_\textrm{Sim}$, two GPS timing receivers were set up to receive the actual and simulated GPS signals, respectively, and generate PPS signals. 
The time difference between the PPS signals from the two receivers is the difference between the actual GPS satellite time and simulated GPS satellite time. 
We developed Python software that recorded the computer's system clock and the difference between the two receivers' timestamps when their PPS outputs were activated on the GPIO pins.

In this manner, $\Delta t_\textrm{Sim}$ was measured every second for 30 min. 
The results are presented in Fig. \ref{fig:TXDelay}.
We used the mean value of the measured $\Delta t_\textrm{Sim}$ as the correction for calibrating the GPS simulator clock. 
As $\Delta t_\textrm{Sim}$ does not significantly vary over time in Fig. \ref{fig:TXDelay}, the residual clock error after the calibration is well within the 50 ms error budget.

\begin{figure}
  \centering
  \includegraphics[width=1.0\linewidth]{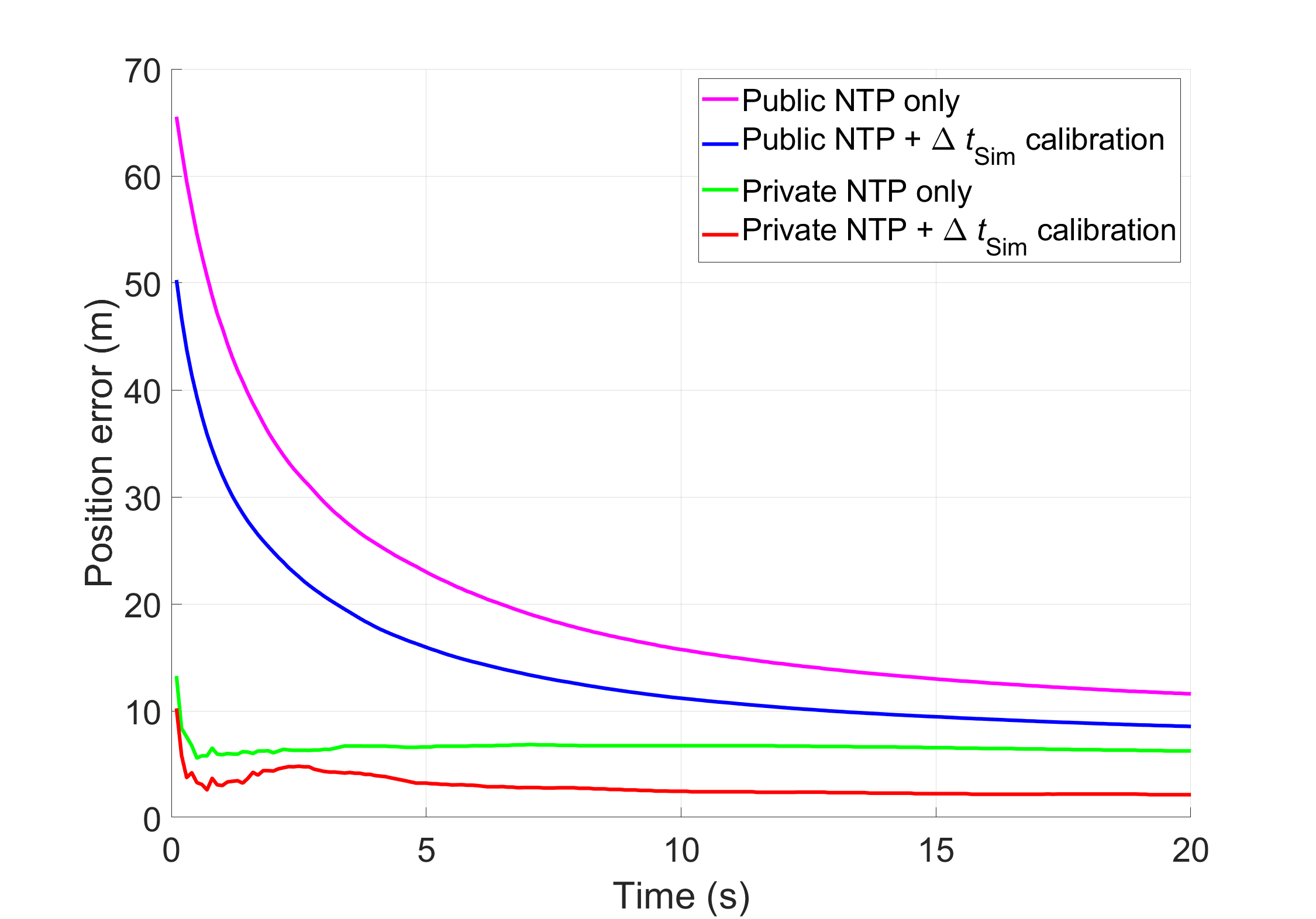}
  \caption{Comparison of the user position errors according to the clock correction methods under a static test.}
  \label{fig:ClockCorrComp}
\end{figure}

\subsection{Positioning performance of the proposed clock synchronization method under a static test} 

To evaluate the improvement of the positioning performance due to the proposed clock correction method, a static test similar to the test in Section \ref{sec:ClockErrReq} was conducted. 
In this test, a GPS user receiver first received the live-sky GPS signals for 30 s. 
Then, the live-sky signal was blocked for 30 s. 
After the blockage, the GPS user receiver received the simulated GPS signals for 20 s, and the accuracy of the position output from the receiver based on the simulated signals was evaluated. 
This scenario considers the outdoor to indoor positioning handover case.
As in Section \ref{sec:ClockErrReq}, the position error was calculated as the difference between the intended user position of the GPS simulator and output position from the user receiver.

Fig. \ref{fig:ClockCorrComp} compares the user position errors according to four clock correction methods.
The position error shortly after receiving the simulated signals is important because a user may not spend a long time within the coverage of a single simulator. 
For example, the car in Fig. \ref{fig:SysOverview} can quickly move from one simulator coverage to the other depending on its velocity. 
The benefit of the private NTP server is evident in Fig. \ref{fig:ClockCorrComp}; it has significantly less position errors than the public NTP server cases for the initial 5 s.
The calibration of the delay in the GPS simulation process, $\Delta t_\textrm{Sim}$, further reduced the position error. 
The maximum position errors and 95\% position errors during the 20-s period are summarized in Table \ref{table:ClockCorrPos}.

In addition to the static test, dynamic tests with a car and a pedestrian were also performed.
The dynamic field test results are presented in Section \ref{sec:Results}.

\section{Optimal Placement of GPS Simulators}
\label{sec:CoverageModel}

\begin{figure}
  \centering
  \includegraphics[width=0.9\linewidth]{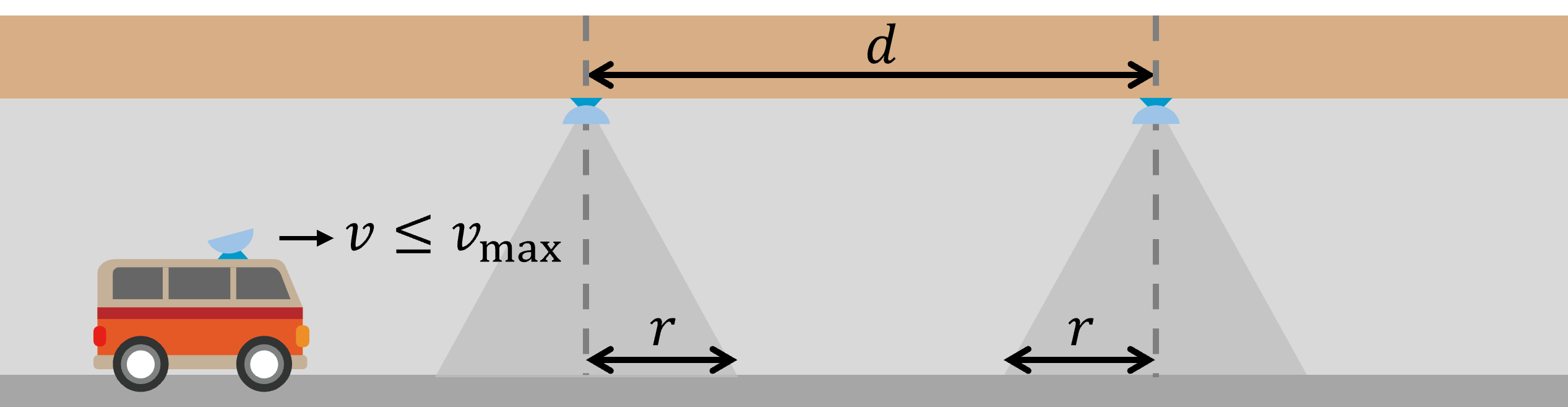}
  \caption{The coverage parameters of the deployed GPS simulators and the velocity of a vehicle in the service area.}
  \label{fig:CoverageModel}
\end{figure}

As mentioned in Section \ref{sec:Intro}, placing a minimum number of GPS simulators for a given area is desirable.
In other words, the separation distance $d$ between the simulators in Fig. \ref{fig:CoverageModel} must be maximized.
Simultaneously, a sufficient number of simulators should be deployed to provide position updates to  users in the given area. 

First, the radius $r$ of the simulator coverage in Fig. \ref{fig:CoverageModel} should be large enough for a user receiver to receive the simulated GPS signals for a period longer than its reacquisition time.
This condition is mathematically expressed as
\begin{equation}
  \label{eqn:ReceptionTime}
  \begin{split}
    t_\mathrm{reacq} &\le t_\mathrm{rcp} \\
    t_\mathrm{rcp} &= \frac{2r}{v} \\
    v &\le v_\textrm{max}
  \end{split}
\end{equation}
where $t_\mathrm{reacq}$ is the reacquisition time of a user receiver after a brief signal loss, $t_\mathrm{rcp}$ is the signal reception time within a simulator coverage, $v$ is the speed of a car, and $v_\textrm{max}$ is the speed limit. 

\begin{figure}
  \centering
  \includegraphics[width=1.0\linewidth]{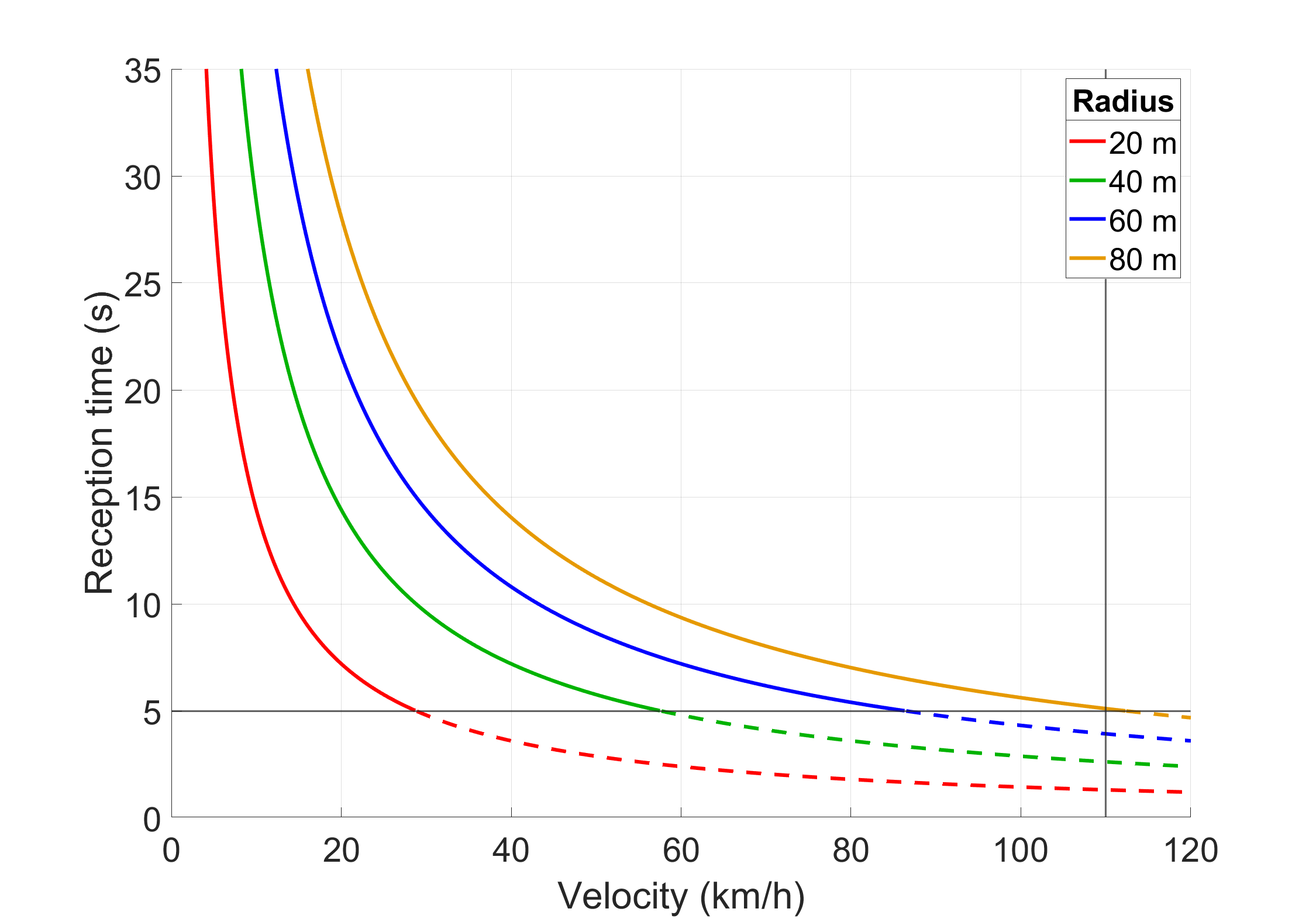}
  \caption{Reception time of the simulated GPS signals vs. vehicle velocity according to the coverage radius. The dotted lines indicate that positioning a GPS receiver may not be possible because the signal reception time is shorter than the reacquisition time of the receiver.}
  \label{fig:ReceptionTime}
\end{figure}

From \eqref{eqn:ReceptionTime}, a $t_\mathrm{rcp}$ vs. $v$ graph can be drawn for a given $r$.
Fig. \ref{fig:ReceptionTime} shows those graphs for four $r$ values when $v_\textrm{max}$ is 110 km/h and $t_\mathrm{reacq}$ is 5 s.
Although the u-blox GNSS receiver utilized for the test in Fig. \ref{fig:ReacqTime} had a reacquisition time of less than 0.5 s under a clock synchronization error of a few hundred ms, the reacquisition performance of a GNSS receiver within a smartphone was significantly worse. 
The observed reacquisition time of a \textit{Samsung Galaxy A5} smartphone under our testbed was approximately 4 s. (The detailed performance of a smartphone under our testbed will be discussed in Section \ref{sec:PhoneTest}.)
To provide a margin, a $t_\mathrm{reacq}$ of 5 s is used in Fig. \ref{fig:ReceptionTime}, which is indicated by the gray horizontal line. 
The gray vertical line represents the speed limit of 110 km/h.  
At the intersection point of two gray lines, we have the following expressions: 
\begin{equation}
  \label{eqn:CoverageRadius}
  \begin{split}
    & t_\mathrm{reacq} = \frac{2r}{v_\mathrm{max}} \\
    & r = \frac{v_\mathrm{max} t_\mathrm{reacq}}{2} 
  \end{split}
\end{equation}

The dotted lines in Fig. \ref{fig:ReceptionTime} imply that a user receiver may not have a position update within the corresponding simulator coverage because a car passes through the coverage earlier than $t_\mathrm{reacq}$.
Therefore, \eqref{eqn:CoverageRadius} and Fig. \ref{fig:ReceptionTime} show that $r \ge \frac{v_\mathrm{max} t_\mathrm{reacq}}{2} = 76.4$ m is required for a position update for a smartphone in a car with a velocity within the speed limit of 110 km/h. 
Thus, the $r$ of 80 m is selected in our case study. 
The coverage radius $r$ can be adjusted by changing the transmitter power of the GPS simulator and gain pattern of the transmitter antenna.

Once $r$ is determined, the separation distance $d$ between simulators in Fig. \ref{fig:CoverageModel} should also be determined. 
If the signal blockage time $t_\mathrm{blk}$ to a user between two adjacent simulator coverages is longer than a certain time limit $t_\mathrm{max}$, fast reacquisition may not be possible because a typical GPS receiver broadens its search space in this situation.  
If it happens, a longer acquisition time $t_\mathrm{acq}$ than $t_\mathrm{reacq}$ is needed to acquire the lost signal. 
The $t_\mathrm{acq}$ of a GPS receiver is usually 30 s. 

These two conditions for a user receiver to have a position update under the proposed indoor positioning system are mathematically expressed as follows:
\begin{equation}
  \label{eqn:BlockageTime}
  \begin{split}
     & t_\mathrm{blk} \le t_\mathrm{max} 
     \;\; \text{or} \;\; 
     t_\mathrm{acq} \le t_\mathrm{rcp} \\
     & t_\mathrm{blk} = \frac{d-2r}{v}
  \end{split}
\end{equation}
Since $t_\mathrm{rcp} = \frac{2r}{v}$, the inequality constraints are represented as follows.
\begin{equation}
  \label{eqn:BlockageTime2}
     t_\mathrm{blk} \le t_\mathrm{max} 
     \;\; \text{or} \;\; 
     v \le \frac{2r}{t_\mathrm{acq}}
\end{equation}

\begin{figure}
  \centering
  \includegraphics[width=1.0\linewidth]{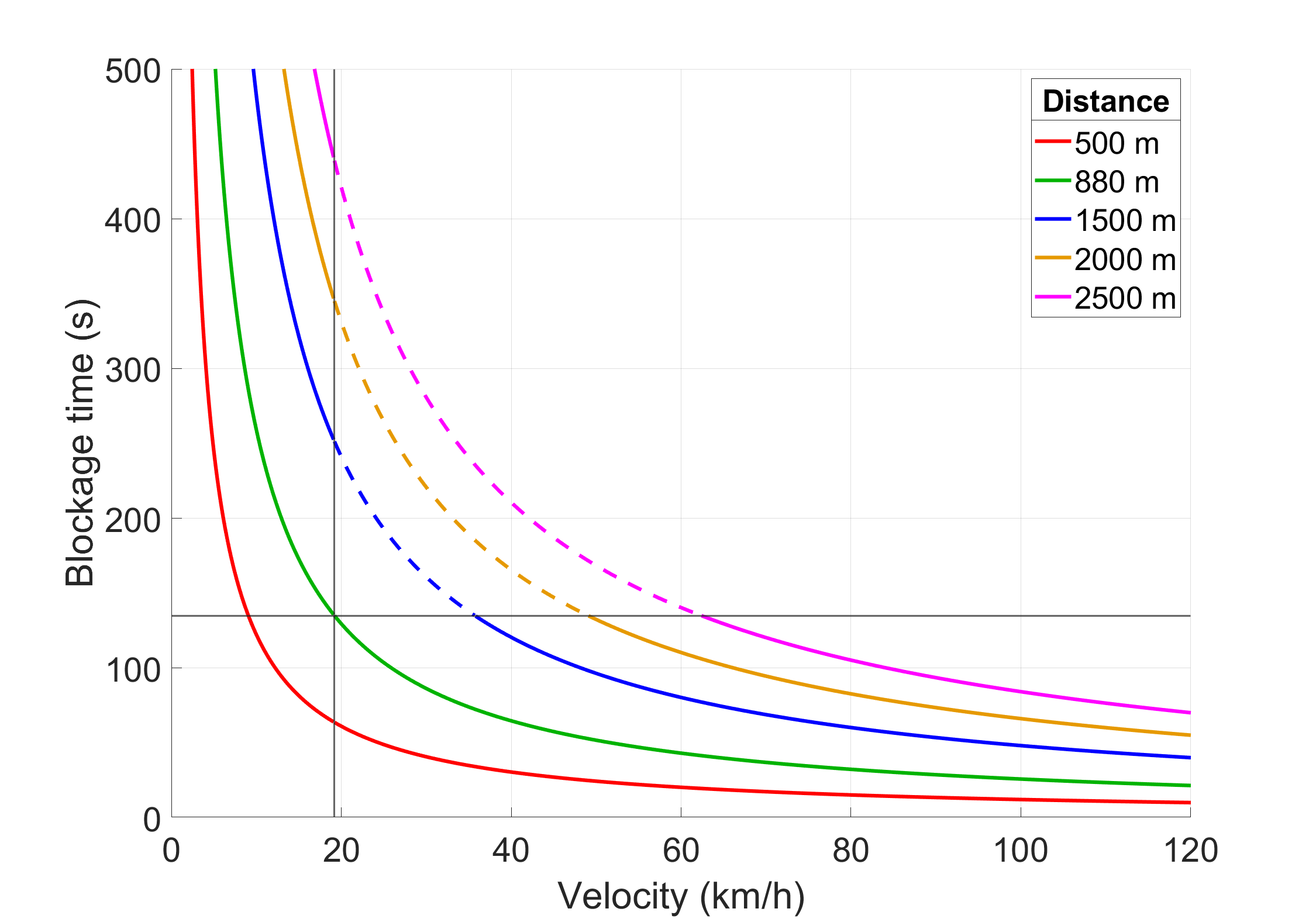}
  \caption{Blockage time vs. vehicle velocity according to the separation distance. The dotted lines indicate that positioning a GPS receiver may not be possible because the blockage time is longer than the time limit for fast reacquisition, and the velocity is too high for the receiver to receive the signals for a sufficient acquisition time.}
  \label{fig:BlockageTime}
\end{figure}

Now we can draw a $t_\mathrm{blk}$ vs. $v$ graph for given $r$ and $d$ using \eqref{eqn:BlockageTime} and \eqref{eqn:BlockageTime2}. 
The graphs for the $r$ of 80 m are presented in Fig. \ref{fig:BlockageTime} for five different $d$ values.
The gray horizontal and vertical lines indicate $t_\mathrm{max}$ and $\frac{2r}{t_\mathrm{acq}}$, respectively. 
According to our smartphone experiment, $t_\mathrm{max}$ was approximately 180 s, and we used 135 s for Fig. \ref{fig:BlockageTime} to be conservative because a different smartphone may have a different $t_\mathrm{max}$. 
A smaller $t_\mathrm{max}$ is more conservative because it reduces the acceptable region with solid lines in Fig. \ref{fig:BlockageTime}.
With the $r$ of 80 m and $t_\mathrm{acq}$ of 30 s, $\frac{2r}{t_\mathrm{acq}}$ is 19.2 km/h (5.3 m/s). 

At the intersection point of two gray lines, the following equations hold:
\begin{equation}
  \label{eqn:CoverageDist}
  \begin{split}
    & t_\mathrm{max} = \frac{d - 2r}{\frac{2r}{t_\mathrm{acq}}} \\
    & d = 2r \left( 1 + \frac{t_\mathrm{max}}{t_\mathrm{acq}} \right)
  \end{split}
\end{equation}
As the solid lines in Fig. \ref{fig:BlockageTime} satisfy the constraints of \eqref{eqn:BlockageTime2}, we know from Fig. \ref{fig:BlockageTime} and \eqref{eqn:CoverageDist} that $d \le 2r \left( 1 + \frac{t_\mathrm{max}}{t_\mathrm{acq}} \right) = 880$ m is necessary for a smartphone in a car to have a position update. 

\begin{table*}
  \caption{Receiver specifications and maximum velocity set for the validation of the proposed coverage model}
  \label{table:CoverageValid}
  \centering
  \begin{tabular}{|c|c|c|c|} 
  \hline
  \makecell{Speed limit [$v_\mathrm{max}$]} & 
  \makecell{Reacquisition time after \\ a brief signal loss [$t_\mathrm{reacq}$]} & 
  \makecell{Maximum allowed signal loss duration \\ for fast reacquisition [$t_\mathrm{max}$]} & 
  \makecell{Acquisition time after a signal loss \\ of $t_\mathrm{max}$ or longer [$t_\mathrm{acq}$]} \\
  \hline
  110 km/h (31 m/s) & 5 s & 135 s & 30 s \\ 
  \hline
  \end{tabular}
\end{table*}

To summarize, the constraints of $r$ and $d$ for the deployment of GPS simulator-based indoor positioning systems are expressed as follows:
\begin{equation}
  \label{eqn:r}
    r \ge \frac{v_\mathrm{max} t_\mathrm{reacq}}{2} 
\end{equation}
\begin{equation}
  \label{eqn:d}
    d \le 2r \left( 1 + \frac{t_\mathrm{max}}{t_\mathrm{acq}} \right)
\end{equation}
The speed limit $v_\mathrm{max}$ is specified by a law, and $t_\mathrm{reacq}$, $t_\mathrm{max}$, and $t_\mathrm{acq}$ are receiver characteristics. 
The specific numbers of those parameters for our case study in this subsection are summarized in Table \ref{table:CoverageValid}. 

When deploying the GPS simulators for the indoor positioning purpose, the power consumption of the simulator and the position update rate for a user should also be considered along with the constraints in \eqref{eqn:r} and \eqref{eqn:d}.
A larger $r$ requires a higher transmission power of the simulator, which causes more power consumption. 
If the simulator with wireless clock synchronization is battery powered, the smallest $r$ that satisfies the constraint of \eqref{eqn:r} is desired to reduce power consumption. 
After $r$ is determined, the largest $d$ that satisfies the constraint of \eqref{eqn:d} can be selected to reduce the number of deployed simulators. 
However, a position update of a user is not frequent if $d$ is large.

If updating user position every 500 m (that is, $d = 500$ m) is more important than reducing the number of simulators, for example, $r \ge \frac{d}{2 \left( 1 + \frac{t_\mathrm{max}}{t_\mathrm{acq}} \right)} = 45.5$ m from \eqref{eqn:d} when the parameters of Table \ref{table:CoverageValid} are used.
In addition, $r \ge \frac{v_\mathrm{max} t_\mathrm{reacq}}{2} = 76.4$ m from \eqref{eqn:r} should also be satisfied. 
Thus, $r = 80$ m and $d = 500$ m can be chosen as an acceptable design for system deployment.

\section{Field Test Results} 
\label{sec:Results}

To evaluate the indoor positioning performance of the proposed system, we implemented an underground testbed with three GPS simulators for field tests.
During the field tests, moving car and walking pedestrian scenarios were considered. 
We utilized a conventional GPS receiver mounted in a car or a smartphone carried by a pedestrian as the user receiver under each test.  



\subsection{Testbed Implementation}

\begin{figure}
  \centering
  \includegraphics[width=1.0\linewidth]{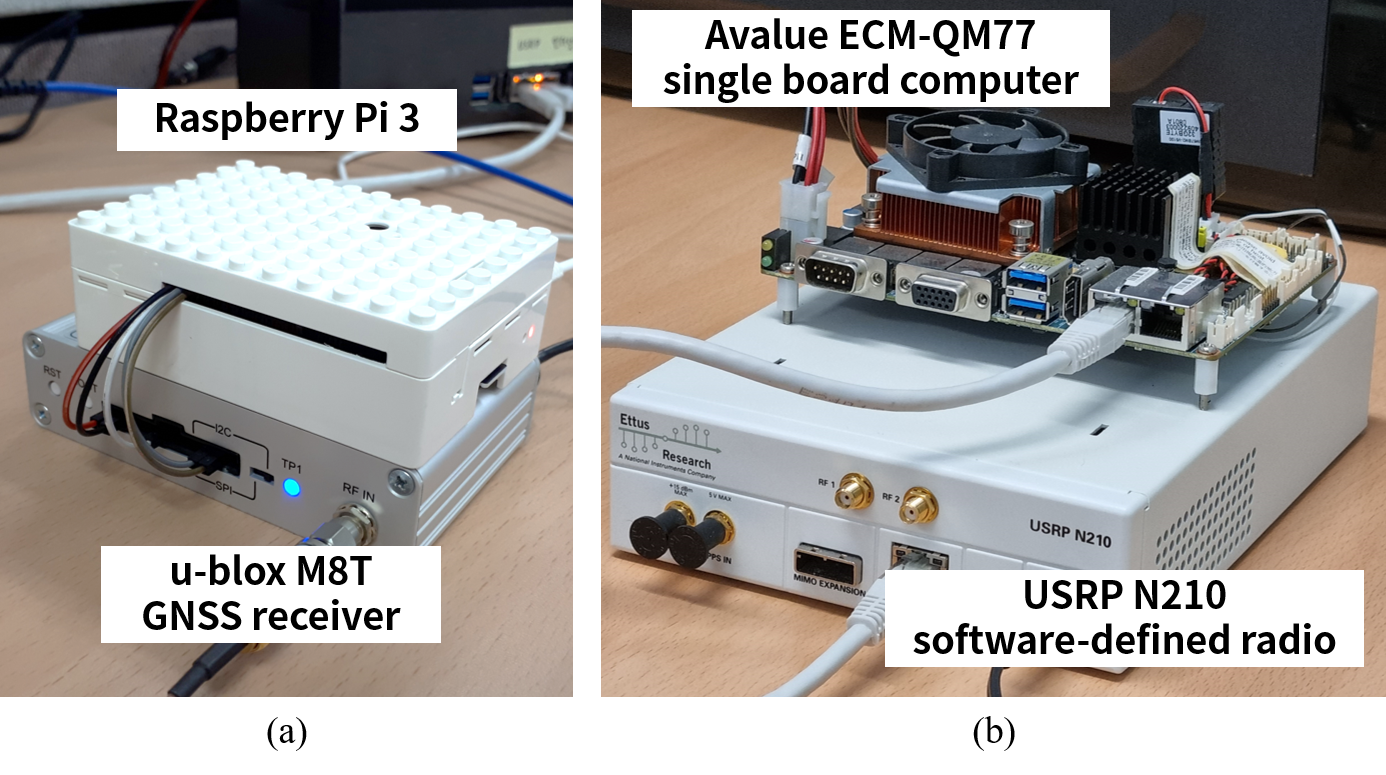}
  \caption{Implemented (a) reference server and (b) GPS simulator for the underground testbed.}
  \label{fig:DevSetup}
\end{figure}

For the underground testbed, the reference server and three GPS simulators were configured as shown in Fig. \ref{fig:DevSetup}.
The GNSS receiver attached to the reference server not only served as a time source for private NTP but also provided real-time navigation data to the three GPS simulators via a wireless LTE network.
As the GPS simulator, an SBC with an Intel i5 mobile processor, \textit{Avalue ECM-QM77}, and a software-defined radio (SDR) platform, \textit{USRP N210}, were utilized.
The SBC simulated GPS signals, and SDR transmitted the simulated signals to users within its coverage. 



The transmission of RF signals in the GPS frequency band is generally prohibited. 
Thus, we obtained the license for experimental radio stations from the Central Radio Management Service of Korea to transmit simulated GPS signals within the underground testbed at Yonsei University in Incheon, Korea.
The license numbers for the three GPS simulators are 91-2022-10-0000013, 91-2022-10-0000014, and 91-2022-10-0000015. 

\begin{figure}
  \centering
  \includegraphics[width=0.9\linewidth]{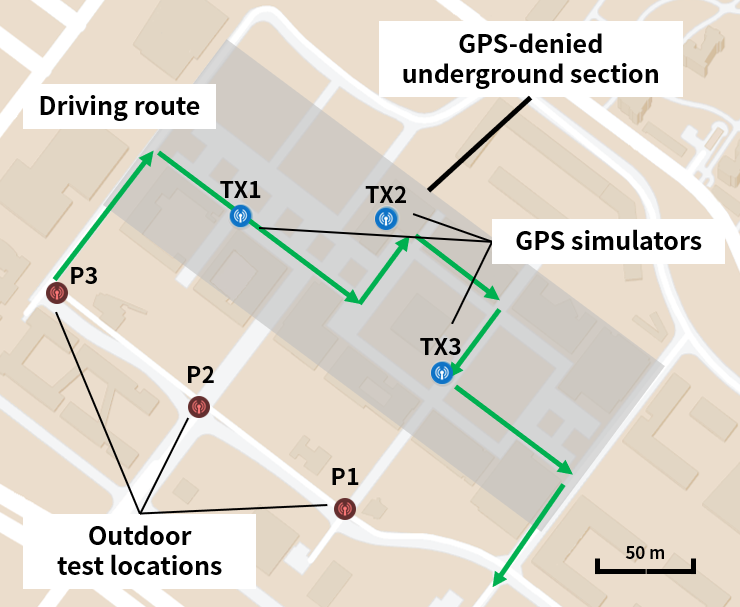}
  \caption{Map of the testbed area at Yonsei University in Incheon, Korea. The driving route, the locations of three GPS simulators in the underground section, and the locations of three outdoor test points are indicated.}
  \label{fig:DrivingTestMap}
\end{figure}

\subsection{Positioning accuracy of a GPS receiver mounted in a car}
\label{sec:DrivingTest}

The driving route for the field test was set as in Fig. \ref{fig:DrivingTestMap}. The route starts from an outdoor location, continues to an underground section, and ends at an outdoor location for testing the capability of indoor and outdoor positioning handover. 
The figure also shows the locations of three GPS simulators.

\begin{figure}
    \centering
    \includegraphics[width=0.9\linewidth]{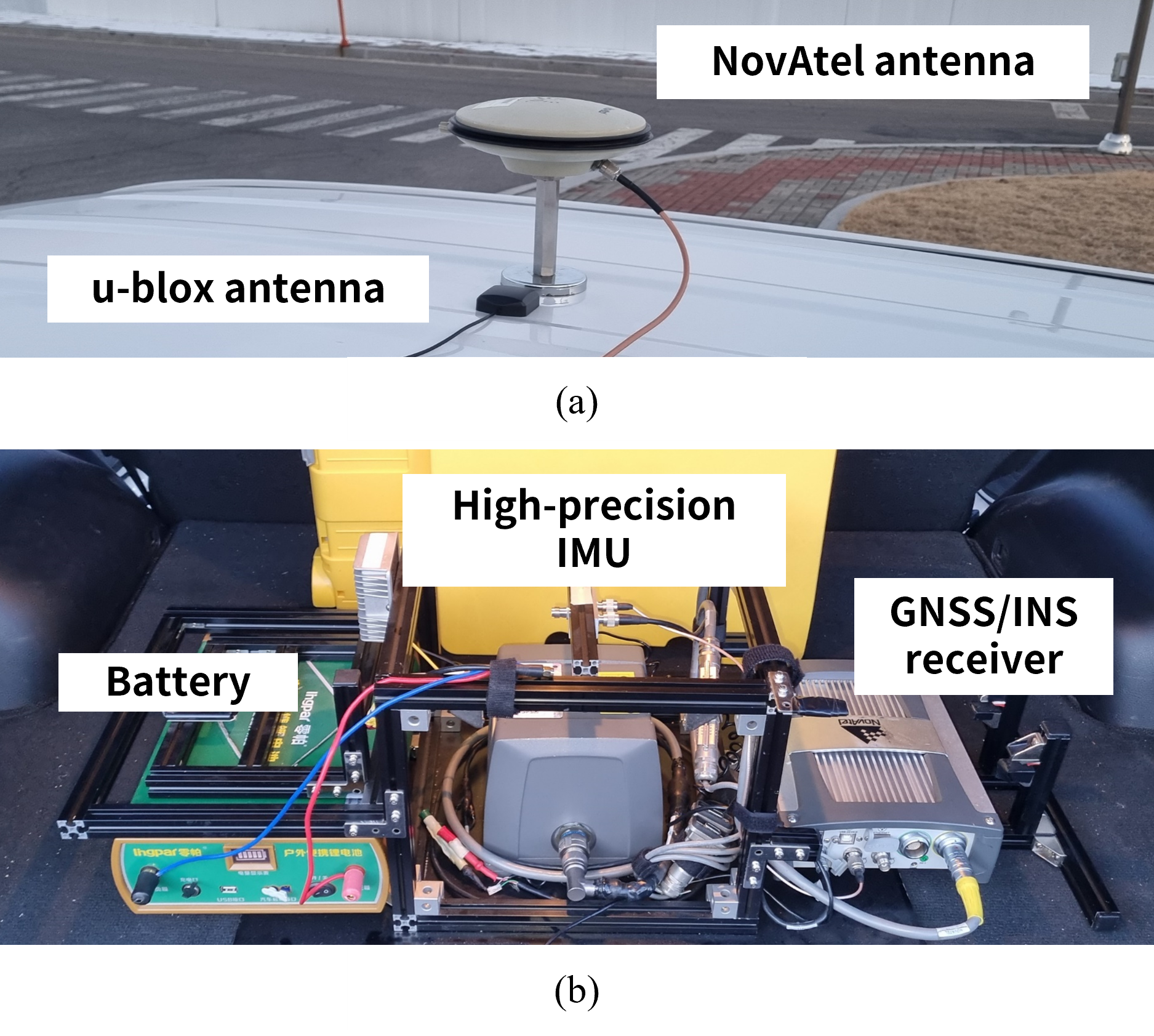}
    \caption{(a) The antennas used in the field test and (b) the GNSS/INS system for obtaining the coordinates of underground car trajectory. The NovAtel and u-blox antennas were connected to the GNSS/INS receiver and the GPS user receiver, which is not displayed in this figure, respectively.}
    \label{fig:TestSetup}
\end{figure}

To specify the intended user position of each simulator in the configuration of Fig. \ref{fig:SysOverview}, the coordinates of the center point of each simulator coverage must be obtained. 
To obtain the coordinates underground, we used a commercial GNSS/INS system as shown in Fig. \ref{fig:TestSetup}.
A NovAtel GNSS/INS receiver, \textit{SPAN-SE}, connected to a high-precision IMU, \textit{UIMU-H58}, was mounted on the vehicle. 
In the outdoor section, the IMU of the GNSS/INS system was calibrated using live-sky GNSS signals.
While driving in the underground section, the position outputs from the GNSS/INS system relied on the IMU because GNSS signals were not available. 
The coordinates of the intended user position of each simulator was determined based on the GNSS/INS trajectories collected in advance. 

We used the \textit{u-blox M8T} receiver as the user receiver for the driving test. 
The positioning accuracy of the u-blox receiver in the testbed was evaluated according to the three clock correction methods for the GPS simulators, which are based on the public NTP server and private NTP server without and with the $\Delta t_\textrm{Sim}$ calibration.

\begin{figure*}
  \centering
  \includegraphics[width=1.0\linewidth]{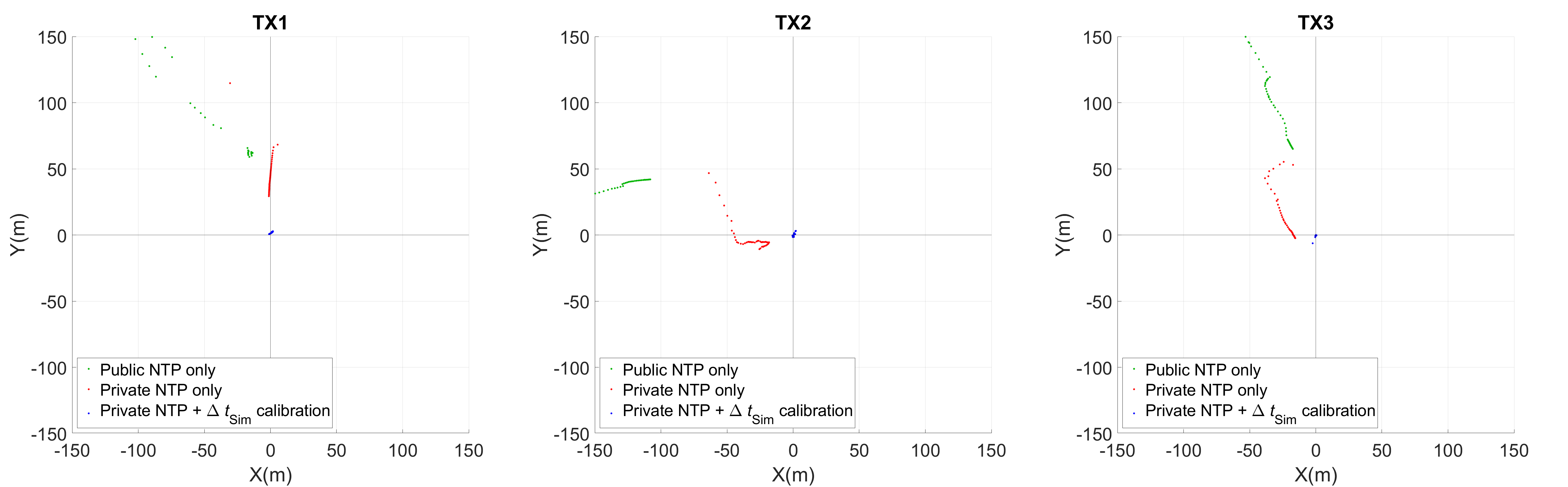}
  \caption{Position outputs from the user receiver during the driving test within the coverages of three simulators (TX1, TX2, and TX3). The origin represents the intended user position of the corresponding simulator. Three clock synchronization methods were applied and compared.} 
  \label{fig:PosResult}
\end{figure*}

The position outputs from the user receiver within each simulator coverage during the driving test are presented in Fig. \ref{fig:PosResult}.
The origin of each plot represents the intended user position of the corresponding simulator (that is, TX1, TX2, or TX3). 
It is clearly shown that the proposed clock synchronization method with the private NTP server and $\Delta t_\textrm{Sim}$ calibration provided the smallest position error. 

\begin{figure*}
  \centering
  \includegraphics[width=1.0\linewidth]{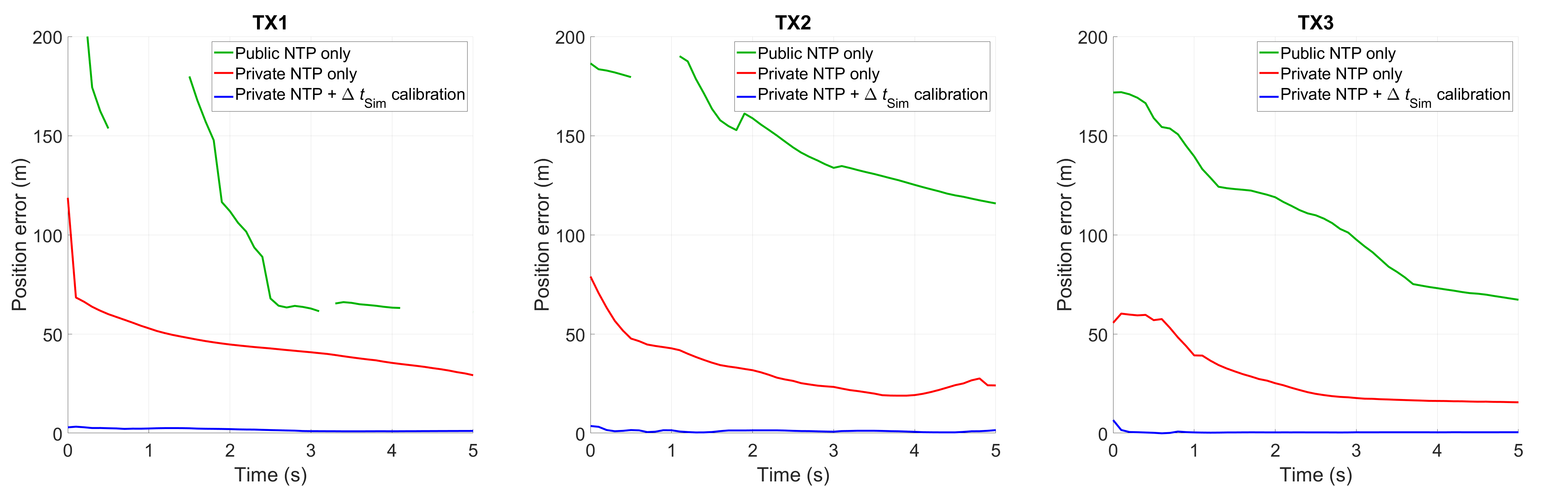}
  \caption{Position outputs from the user receiver during the driving test within the coverages of three simulators (TX1, TX2, and TX3). The \textit{x}-axis represents the time after receiving the simulated signals from the corresponding simulator. Three clock synchronization methods were applied and compared.}
  \label{fig:ErrResult}
\end{figure*}

Fig. \ref{fig:ErrResult} shows the user position outputs for the time after receiving the signals from each simulator.
The proposed method with the private NTP server and $\Delta t_\textrm{Sim}$ calibration enabled significantly smaller position errors than the other cases throughout the 5-s period within the coverage of each simulator.
Table \ref{table:DrivingResult} compares the average, standard deviation, and root-mean-square (RMS) values of the user position error during the driving test.
The average position error of the public NTP server or private NTP server without $\Delta t_\textrm{Sim}$ calibration case was 121.8 or 35.7 m, respectively.
In contrast, when the clock synchronization was performed based on the private NTP server and $\Delta t_\textrm{Sim}$ calibration, the average position error was only 3.7 m. 
As the proposed method had a small position error right after receiving simulated signals, the positioning handover between live-sky and simulated signals was successful. 


\begin{table}
  \caption{Comparison of the user position errors according to the clock correction methods during the driving test}
  \label{table:DrivingResult}
  \centering
  \begin{tabular}{|c|c|c|c|} 
  \hline
  \makecell{} & 
  \makecell{Average} & 
  \makecell{Standard  \\ deviation} & 
  \makecell{RMS} \\
  \hline
  \makecell{Public NTP server only} & 121.8 m & 38.4 m & 128.9 m \\
  \hline
  \makecell{Private NTP server only} & 35.7 m & 14.5 m & 38.6 m \\
  \hline
  \makecell{Private NTP server with \\ $\Delta t_\textrm{Sim}$ calibration} &
  3.7 m & 1.2 m & 4.1 m \\
  \hline
  \end{tabular}
\end{table}

\subsection{Comparison of Positioning Accuracy with the Live-Sky GPS Signal Case}
\label{sec:OutdoorTest}

In addition to the underground test, the positioning accuracy of the proposed system was tested in an outdoor environment. 
Three outdoor test locations are indicated in Fig. \ref{fig:DrivingTestMap}. 
The simulated GPS signals from the proposed system were passed into a GPS receiver via a wired connection during the field test because our license for the experimental radio stations is limited to underground radio transmissions only.
This restriction is to prevent possible interference with live-sky GPS signals for outdoor users. 

\begin{table}
  \caption{Comparison of the user position errors between the live-sky GPS and simulated GPS signal cases at the outdoor test locations}
  \label{table:OutdoorResult}
  \centering
  \begin{tabular}{|c|c|c|c|} 
  \hline
  \makecell{} & 
  \makecell{Average} & 
  \makecell{Standard  \\ deviation} & 
  \makecell{RMS} \\
  \hline
  \makecell{Live-sky GPS signals} & 2.8 m & 0.8 m & 3.0 m \\
  \hline
  \makecell{Simulated GPS signals} & 4.4 m & 1.3 m & 4.8 m \\
  \hline
  \end{tabular}
\end{table}

Table \ref{table:OutdoorResult} shows the average, standard deviation, and RMS values of the user position errors during the 5-s signal reception period at each outdoor test location.
The average outdoor position error of the proposed system was 4.4 m, which is similar to the underground case of 3.7 m in Table \ref{table:DrivingResult}. 
Although the simulated signals provided higher position errors than the 2.8 m error from the live-sky GPS signals, the demonstrated underground and outdoor errors were smaller than the coverage of each simulator. 
Thus, it is confirmed that the proposed system can serve its purpose of providing periodic position updates to users within its coverage.

\subsection{Positioning accuracy of a smartphone carried by a pedestrian} 
\label{sec:PhoneTest}

Along with the underground driving test using a dedicated GPS user receiver and antenna, we conducted tests to evaluate the positioning accuracy of a smartphone carried by a pedestrian in the same testbed.
The pedestrian tests were conducted under two conditions: with and without transmitting simulated GPS signals. 
The proposed clock synchronization method was applied, and a \textit{Samsung Galaxy A5} was utilized as the smartphone under test. 
During the pedestrian tests, the screen of the \textit{Google Maps} application of the smartphone was recorded. 
To prevent receiving possible positioning assistance from other radio signals than GPS signals, the smartphone was set to the airplane mode during the tests.
 

\begin{figure}
  \centering
  \includegraphics[width=0.9\linewidth]{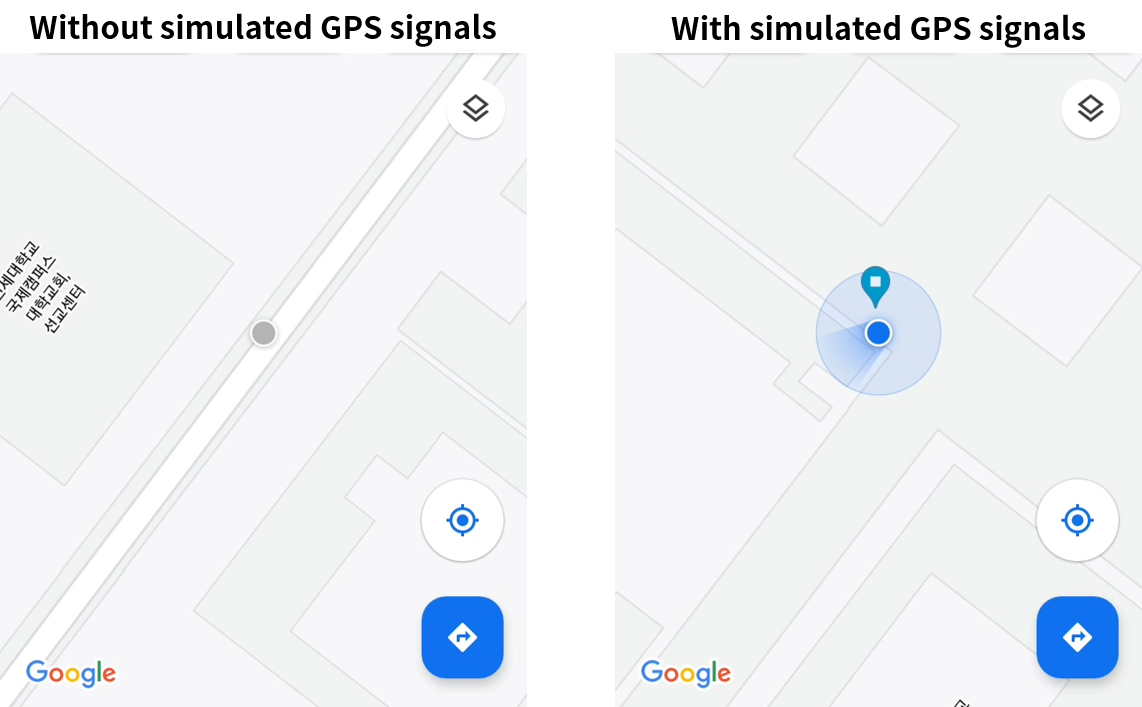}
  \caption{Comparison of the display of a conventional map application of the smartphone under the underground pedestrian tests. The left screen corresponds to the case without the simulated GPS signals, and the right screen is the case with the simulated GPS signals. The underground user position was correctly displayed in the map application when the simulated GPS signals were transmitted.}
  \label{fig:PhoneTest}
\end{figure}

As shown in Fig. \ref{fig:PhoneTest}, when the simulated GPS signals were not transmitted underground, the smartphone could not acquire its location and displayed the last confirmed location as a gray circle.
In contrast, when the simulated GPS signals were transmitted, the smartphone successfully acquired its underground location and displayed it as a blue circle.
The pinned location in the right screen of Fig. \ref{fig:PhoneTest} indicates the intended user position of the corresponding GPS simulator. 
The displayed user location on the map was close to the intended user position. 
The average position error during the pedestrian test with the GPS simulators was 9.6 m.

The position error of the smartphone was larger than the case with a dedicated GPS receiver and antenna mounted on the vehicle. 
This is a typical result because the GNSS chipsets and antennas within smartphones are known to have poor measurement quality \cite{Yi22:Native, Benvenuto22:Adaptive}.
Nevertheless, a smartphone user can obtain the useful position information from the proposed indoor GPS simulators even in the underground areas.

\subsection{Wireless Security Issues} 
\label{sec:Security}

It is important to note that there are potential wireless security issues related to the proposed indoor positioning system.
The GPS simulator of the proposed system can be abused as a GPS spoofer.
To prevent unauthorized manipulations, secure communication protocols between the simulator and reference server should be established.
Furthermore, the intended user position of each simulator should be stored in a read-only memory (ROM) within the simulator to add a layer of defense.

\section{Conclusion}
\label{sec:Conclusion}

The GPS simulator-based indoor positioning system is appropriate for providing periodic position updates to the GPS users within its coverage. 
With this method GPS user receivers do not require any hardware or software modifications, and the number of GPS simulators required to cover a given area is small.
However, deploying such a system can be difficult if the wiring between each simulator and an outdoor GPS antenna is required for clock synchronization.  
To reduce the difficulty and cost of simulator deployment, we proposed methods to synchronize the simulator clocks wirelessly. 
We first analyzed the required synchronization accuracy for seamless indoor and outdoor positioning handover. 
Then, wireless clock synchronization within the required accuracy was achieved using the private NTP server and calibration method. 
Furthermore, the constraints regarding the coverage radius and separation distance of the GPS simulators were derived for optimal system deployment.
Field tests demonstrated the positioning accuracy of the proposed system for a driving car with a GPS receiver and a pedestrian with a smartphone in the underground testbed with three GPS simulators.
To the authors' knowledge, this is the first demonstration of a GPS simulator-based indoor positioning system with wireless clock synchronization.



\bibliographystyle{IEEEtran}
\bibliography{mybibliography, IUS_publications}

\phantomsection

\begin{IEEEbiography}[{\includegraphics[width=1in,height=1.25in,clip,keepaspectratio]{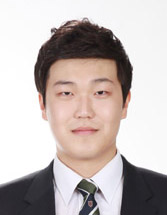}}]{Woohyun Kim} received the B.S degree in integrated technology from Yonsei University, Incheon, Korea, in 2015. He is currently pursuing the Ph.D. degree in integrated technology at Yonsei University, Incheon, Korea. His research interests include complementary navigation systems, indoor navigation, and intelligent unmanned systems. Mr. Kim was a recipient of the Undergraduate and Graduate Fellowships from the ICT Consilience Creative Program supported by the Ministry of Science and ICT, Korea.
\end{IEEEbiography}

\begin{IEEEbiography}[{\includegraphics[width=1in,height=1.25in,clip,keepaspectratio]{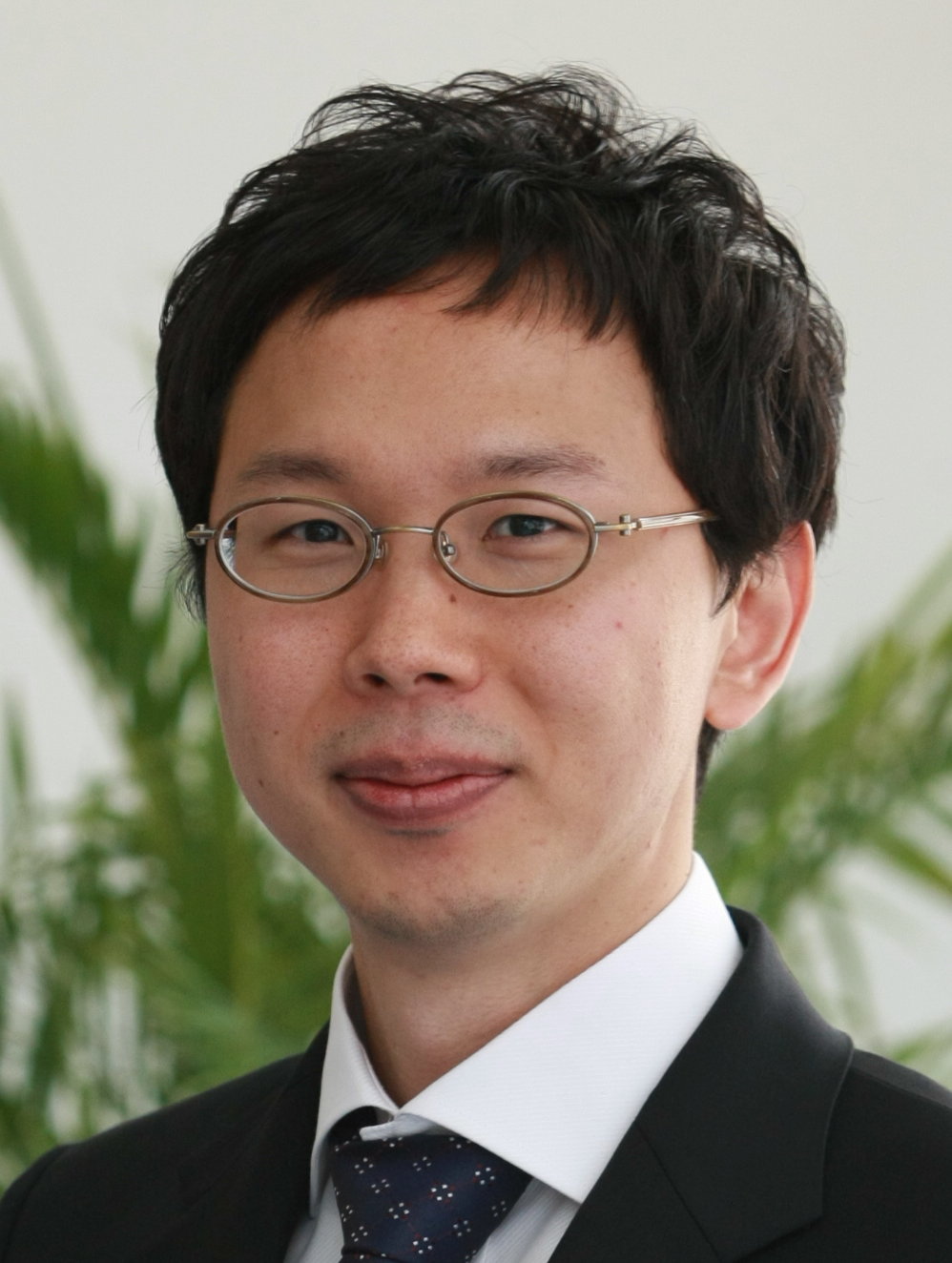}}]{Jiwon Seo} (M'13) received the B.S. degree in mechanical engineering (division of aerospace engineering) in 2002 from Korea Advanced Institute of Science and Technology, Daejeon, Korea, and the M.S. degree in aeronautics and astronautics in 2004, the M.S. degree in electrical engineering in 2008, and the Ph.D. degree in aeronautics and astronautics in 2010 from Stanford University, Stanford, CA, USA. He is currently an associate professor with the School of Integrated Technology, Yonsei University, Incheon, Korea. His research interests include GNSS anti-jamming technologies, complementary PNT systems, and intelligent unmanned systems. 
Prof. Seo is a member of the International Advisory Council of the Resilient Navigation and Timing Foundation, Alexandria, VA, USA, and a member of the Advisory Committee on Defense of the Presidential Advisory Council on Science and Technology, Korea.
\end{IEEEbiography}

\EOD

\vfill

\end{document}